\documentclass[aps,prd,10pt,nofootinbib,twocolumn,eqsecnum,showpacs,showkeys,superscriptaddress,preprintnumbers,altaffilletter]{revtex4-1}

\usepackage{graphicx}
\usepackage{dcolumn}
\usepackage{amssymb}
\usepackage{amsmath}
\usepackage{amsfonts}
\usepackage{amsbsy}
\usepackage{color}
\usepackage{rotating}
\usepackage[english]{babel}
\usepackage{hyperref}
\usepackage[utf8]{inputenc}
\usepackage{epsfig}
\usepackage{soul}

\def\be{\begin{equation}}
\def\ee{\end{equation}}

\def\bea{\begin{eqnarray}}
\def\eea{\end{eqnarray}}

\begin{document}

\title{The \textit{Umami} Chaplygin model}

\author{Ruth Lazkoz}
\email{ruth.lazkoz@ehu.es}
\affiliation{Department of Theoretical Physics, University of the Basque Country UPV/EHU, P.O. Box 644, 48080 Bilbao, Spain}
\author{Mar\'ia Ortiz Ba\~nos}
\email{maria.ortiz@ehu.eus}
\affiliation{Department of Theoretical Physics, University of the Basque Country UPV/EHU, P.O. Box 644, 48080 Bilbao, Spain}
\author{Vincenzo Salzano}
\email{vincenzo.salzano@usz.edu.pl}
\affiliation{Institute of Physics, Faculty of Mathematics and Physics, University of Szczecin, Wielkopolska 15, 70-451 Szczecin, Poland}

\date{\today}

\begin{abstract}
In this work we study in detail a phenomenological generalization of the Chaplygin cosmological model, which we call as \textit{umami} Chaplygin model.
%\enzo{\st{This model includes an early phase dominated by dust-like matter followed by a dark fluid phase at late times which causes the accelerated expansion of the universe.}}
We consider three different cosmological background scenarios in which our fluid can play three different roles: only as a dark energy component; as a dark matter and dark energy component; and as a dark plus baryonic matter and dark energy component. With such analysis we explore the possibility to unify the dark fluids under one single component within the context of General Relativity. We test this hypothesis against the main available data related to the cosmological background, namely: Type Ia supernovae; Baryon Acoustic Oscillations; Quasars; Gamma Ray Bursts; Hubble data from cosmic chronometers; and Cosmic Microwave Background. We eventually compare the statistically efficiency and reliability of our model to describe observational data with respect to the standard $\Lambda$CDM model by means of the Bayesian Evidence. Final results point to a positive (albeit not strong) evidence in favor of a possible unification of dark energy and dark matter with a the \textit{umami} fluid.
\end{abstract}

\maketitle

%%%%%%%%%%%%%%%%%%%%%%%%%%%%%%%%%%%%%%%%%%%%
\section{Introduction} \label{sec:intro}
One of the most studied problems nowadays in cosmology is the nature of the accelerated expansion of the Universe at late times. Since the late 20th century \cite{Riess:1998cb,Perlmutter:1998np,Perlmutter:2003kf}, this phenomenon has been reported and confirmed by several different observational probes \cite{Aghanim:2018eyx,sdss1,sdss2}. In contrast with the clear signatures we have about it, we are still far from a possible agreement concerning its theoretical explanation.

It is very well-known and established nowadays that most of the work can bifurcate into two branches. One points toward the modification of the Hilbert-Einstein action, which drives into a generalization of Einstein's field equations \cite{Berti:2015itd, Review}. The other line of research attempts to modify the energy density part of such equations by introducing a new cosmological fluid generally dubbed as dark energy (DE). DE has been extensively studied both theoretically \cite{Frieman:2008sn,Li:2011sd} and phenomenologically, with the most common approach consisting in replacing the standard cosmological constant with another fluid whose equation of state is not constant, but varying on time, and depending on a variety of extra parameters  \cite{Cooray:1999da,Efstathiou:1999tm,Astier:2000as,Goliath:2001af,Chevallier,Linder,Weller:2001gf,Sahni:2002fz,Padmanabhan:2002vv,Wetterich:2004pv,Jassal:2004ej,Komatsu:2008hk,Wang:2008vja,Ma:2011nc,Barboza:2011py,Sendra:2011pt}.
In this work we will focus on this line. 

As the standard model $\Lambda$CDM \cite{Carroll:1991mt,Sahni:1999gb,Peebles:2002gy} works relatively fine \cite{Planck,Aubourg:2014yra} despite of its theoretical and observational issues \cite{beyond,Bernal2016}, if one wants to consider a background extra fluid, the new model will have, at least, to satisfy two limits which preserve the the expected expansion rate at high and low redshift, that is, a matter dominated epoch at early universe and an accelerated expansion epoch at late times.

The model we have chosen belongs to the so-called Chaplygin-like models. These arose originally as an alternative to quintessence, the latter describing a new type of matter which is represented by a scalar field and whose behaviour describes the transition from a universe filled with dust to an exponentially expanding universe. Chaplygin-like models are characterized by describing this transition in terms of a single perfect fluid with an exotic equation of state. This means that we have dark matter and dark energy unified in a single fluid \cite{Kamenshchik:2001cp}. 
%From the phenomenological point of view, the fact of having the whole dark component unified is an interesting feature \maria{\ul{because, in order to study the evolution of the background, what we need it is a parameterization which mimics the effects we observe. 
Moreover, the unified description of dark energy and matter has an interesting characteristic as it can be reinterpreted as a solution to an interacting matter plus vacuum model \cite{DeSantiago:2012xh}.

Many generalizations of this fluid have been proposed since then, the most studied in the literature are modified Chaplygin, extended Chaplygin and generalized Chaplygin. See \cite{Pourhassan:2014ika} for a brief review. The interesting feature of the specific model we present here is that our non-linear equation of state at late times does not have a de Sitter universe as limit case but a fluid with a constant equation of state $w$. Apart from the Chaplygin-like models, non-linear barotropic equations of state also appear in other works, the most recent one being \cite{Berteaud:2018ifl}. They focus on different descriptions, as quadratic models to represent dark energy and unified dark matter \cite{Ananda:2005xp}, non-linear EoS for phantom fluids \cite{Dutta:2010yu} or the Born-Infeld type fluid model \cite{Chen:2015kza}.

We will treat our fluid in a polyvalent way in order to extract as much information as we can. Firstly we will consider the possibility that it acts as the usual dark energy component. Secondly, as a dark matter-dark energy unified component and finally, as the full (i.e. dark and baryonic) matter and dark energy content. The two last cases are the most intriguing ones because in that case we realize the idea of having all the ``dark'' behaviour unified into a single fluid. The work is organized as follows: We begin with Sec.~\ref{sec:introducing}  by introducing our model as well as the different scenarios we are interested in; in Sec.~\ref{sec:data} we describe the observational data used in our analysis and in Sec.~\ref{sec:results} we discuss the obtained results and give some remarkable conclusions.

\section{Model}\label{sec:introducing}

We study what we have named the \textit{umami} Chaplygin model described by the following equation of state:
\be\label{eq}
P=-\frac{\rho}{\displaystyle{\frac{1}{|w|}}+\displaystyle{\frac{\rho^2}{|A|}}}
\ee
where $P$ and $\rho$ are the pressure and the energy density of our fluid and $w$ and $A$ are real constants. Let us note that, on one hand, at high energy density pressures the fluid behaves as
\be
P\rightarrow-\frac{|A|}{\rho},
\ee
which is the equation of state of the original form of the Chaplygin fluid. On the other hand, the low energy density limit gives
\be
P\rightarrow-|w|\rho,
\ee
which is the equation of state of a perfect fluid with negative effective EoS parameter; thus, the fluid will be playing the role of some dark energy component.

We assume the fluid independently fulfills the usual conservation equation,
\be\label{conti}
\dot{\rho}+3H\left(\rho+P\right)=0,
\ee
where $\cdot\equiv d/dt$ and $H=\frac{\dot{a}}{a}$ is the Hubble parameter. This can be solved analytically thus obtaining the following implicit expression for $\rho$:
\bea\nonumber
&&\rho^{\frac{1}{1-|w|}}\Big||A|-|A|\,|w|+|w|\,\rho^2 \Big|^{\frac{|w|}{2(|w|-1)}}=\\\label{rho}
&&\rho_0^{\frac{1}{1-|w|}}\Big||A|-|A|\,|w|+|w|\,\rho_0^2 \Big|^{\frac{|w|}{2(|w|-1)}}a^{-3},
\eea
where the subindex $0$ means that the quantity is evaluated at the present time.

To set up the background cosmological evolution we have solved Eq.~(\ref{rho}) numerically in order to find $\rho$. Given the non-linearity of the solution and for practical reasons, we have changed the variable $\rho$ to the dimensionless matter parameter $\overline{\Omega}_f\equiv\frac{\rho}{\rho_{c,0}}$, where $\rho_{c,0} = \frac{3 H^{2}_{0}}{8\pi G}$ is the critical density of the Universe at the present time. Note that this newly defined $\overline{\Omega}_{f}$ parameter is not fully equivalent to the standard dimensionless density parameter $\Omega_{f}$, which should have been defined as $\rho/\rho_{c}$. The two versions are equivalent only in the present-time limit, i.e. $\overline{\Omega}_{f,0} = \Omega_{f,0}$, which is the case we are mostly interested in, if we want to compare our results with literature. Thus, Eq.~({\ref{rho}}) will become:
\begin{eqnarray}\label{rho_to_omega}
\overline{\Omega}_{f}^{\frac{1}{1-|w|}}\left||\overline{A}|-|\overline{A}|\,|w|+|w|\,\overline{\Omega}_{f}^2\right|^{\frac{|w|}{2(|w|-1)}}= \nonumber \\ \overline{\Omega}_{f,0}^{\frac{1}{1-|w|}}\left||\overline{A}|-|\overline{A}|\,|w|+|w|\,\overline{\Omega}_{f,0}^2\right|^{\frac{|w|}{2(|w|-1)}} a^{-3}\; ,
\end{eqnarray}
where also the parameter $A$ has been redefined as $\overline{A} \equiv A/\rho^{2}_{c,0}$. The solution to Eq.~(\ref{rho_to_omega}) allows us to add the \textit{umami} fluid in a more direct observationally-related form into the expression of the Friedmann equation obtained after the standard assumption of a FLRW metric background.

In particular, we have studied three different scenarios, where the \textit{umami} fluid plays different multiple roles:
\bea\nonumber\label{eq:scenarios}
E_1(z)&=&\sqrt{\Omega_m (1+z)^{3}+\Omega_r (1+z)^{4}+\overline{\Omega}_f(z)+\Omega_k (1+z)^{2}}\\
\\\nonumber
E_2(z)&=&\sqrt{\Omega_b (1+z)^{3}+\Omega_r (1+z)^{4}+\overline{\Omega}_f(z)+\Omega_k (1+z)^{2}}\\
\\
E_3(z)&=&\sqrt{\Omega_r (1+z)^{4}+\overline{\Omega}_f(z)+\Omega_k (1+z)^{2}}
\eea
where $E(z)\equiv H(z)/H_0$ is the dimensionless Hubble parameter, and $\Omega_m, \Omega_b, \Omega_r$ are the normalized densities of total matter (baryons and dark matter), baryons-only and radiation today, while $\Omega_k\equiv k / H^2_0$ corresponds to the spatial curvature. In the first scenario our fluid $\overline{\Omega}_f$ represents only a dark energy component; in the second one, we unify the dark matter and dark energy content into $\overline{\Omega}_f$; finally, in the third scenario, $\overline{\Omega}_f$ would include the total matter content and the dark energy.

We have to take special attention to the two latter models: in principle, we have no clue about how much of our \textit{umami} fluid should behave as dark energy and/or dark matter. We are giving here a phenomenological proposal, with no physical insight about its possible physical origin. For that, in case $2$, we define the parameter $f_{\Omega_{DM}}$ which is the  fraction of $\overline{\Omega}_f$ that is dark matter. In this way we have that the total matter content is $\Omega_m=\Omega_b+f_{\Omega_{DM}}\cdot\overline{\Omega}_f$. In model $3$, something similar happens. We need to know which part of our fluid is dark matter and which part is baryonic matter. In this case we have $\Omega_m=(f_{\Omega_{b}}+f_{\Omega_{DM}})\cdot\overline{\Omega}_f$, where $f_{\Omega_b}$ is the fraction of baryons in $\overline{\Omega}_f$.

In order to describe numerically the general function $\overline{\Omega}_f(a)$ (where $a$ is the scale factor) and save time when performing the statistical data analysis, we have built a grid on the parameters $\{a,\overline{\Omega}_{f,0},w,\overline{A}\}$ and defined the final \textit{umami} density function as an interpolating function on this grid. The chosen grid is:
\bea
1\cdot 10^{-11}&<a&<1\\
0.3&<\overline{\Omega}_{f,0}&<1.5\\
%-0.9999&<w&<-0.015\\
-1&<w&<-0.015\\
-500000&<\overline{A}&<-0.025
\eea
We want to stress that we keep $w>-1$ in order to avoid phantom fields and the ``singular'' value $w=-1$. The range of $A$ is large enough to cover all the possible physically meaningful values for our model; but we need to specify also that the grid is not globally uniform. That is because as well as for $w\leq -1$ and for some combination of $w> -1$ values with some ranges of the $A$ parameter, the numerical solution of Eq.~(\ref{rho}) may not be physical or not univocal. In fact, for some combinations we might have negative densities or two positive solutions with one branch leading to increasing density in time. While this could be considered meaningful in some singularity scenarios \cite{Dabrowski:2014fha}, we have avoided them and stuck to a more standard evolution of the cosmological background.

\section{Data}\label{sec:data}

We use a combination of various current observational data to constrain the Umami Chaplygin models described previously. In this section, we describe the
cosmological observations used in this work. We will only consider the observational data related to the expansion history of the universe, i.e., those describing the distance-redshift relations. Specifically, we use Type Ia Supernovae (SNeIa), quasars, Gamma Ray Bursts (GRB), cosmic microwave background (CMB) distance priors, Baryon Acoustic Oscillations (BAO) data and the expansion rate data from early-type galaxies (ETG).

\subsection{Hubble data}

We use a compilation of Hubble parameter measurements estimated by the differential evolution of passively evolving early-type galaxies used as cosmic chronometers, in the redshift range $0<z<1.97$, and recently updated in \cite{Moresco15}. The corresponding $\chi^2_{H}$ estimator is defined as
\begin{equation}\label{eq:hubble_data}
\chi^2_{H}= \sum_{i=1}^{24} \frac{\left( H(z_{i},\boldsymbol{\theta})-H_{obs}(z_{i}) \right)^{2}}{\sigma^2_{H}(z_{i})} \; ,
\end{equation}
with $\sigma_{H}(z_{i})$ the observational errors on the measured values $H_{obs}(z_{i})$, $\boldsymbol{\theta}$ the vector of the cosmological background parameters.

\subsection{Pantheon Supernovae data}

We used the SNeIa data from the Pantheon compilation \cite{Scolnic:2017caz}.
This set is made of $1048$ SNe covering the redshift range $0.01<z<2.26$. The $\chi^2$ in this case is defined as
\begin{equation}
\chi^2_{SN} = \Delta \boldsymbol{\mathcal{F}}^{SN} \; \cdot \; \mathbf{C}^{-1}_{SN} \; \cdot \; \Delta  \boldsymbol{\mathcal{F}}^{SN} \; ,
\end{equation}
with $\Delta\boldsymbol{\mathcal{F}} = \mathcal{F}_{theo} - \mathcal{F}_{obs}$, the difference between the observed and the theoretical value of the observable quantity for SNeIa, the distance modulus; and $\mathbf{C}_{SN}$ the total covariance matrix
%(for a discussion about all the terms involved in its derivation, see \cite{Betoule}).
The predicted distance modulus of the SNe, $\mu$, given the cosmological model, is defined as
\begin{equation}\label{eq:m_jla}
\mu(z,\boldsymbol{\theta}) = 5 \log_{10} [ d_{L}(z, \boldsymbol{\theta}) ] +\mu_0 \; ,
\end{equation}
where $D_{L}$ is the dimensionless luminosity distance given by
\be
d_L(z,\theta_c)=(1+z)\int_{0}^{z}\frac{dz'}{E(z')}
\ee
with $E(z)=H(z)/H_0$. However the previous expression of $\chi^2_{SN}$ would contain the nuisance parameter $\mu_0$, which depends on the Hubble constant, the speed of light $c$ and the SNeIa absolute magnitude. In order to get rid of the degeneracy intrinsic to the definition of the parameters, we marginalize analytically over $\mu_0$. All details can be found in \cite{conley}. Finally, the resulting $\chi^2$ gives
\be\label{eq:chis}
\chi^2_{SN}=a+\log \frac{d}{2\pi}-\frac{b^2}{d},
\ee
where $a\equiv\left(\Delta \boldsymbol{\mathcal{F}}_{SN}\right)^T \; \cdot \; \mathbf{C}^{-1}_{SN} \; \cdot \; \Delta  \boldsymbol{\mathcal{F}}_{SN}$, $b\equiv\left(\Delta \boldsymbol{\mathcal{F}}^{SN}\right)^T \; \cdot \; \mathbf{C}^{-1}_{SN} \; \cdot \; \boldsymbol{1}$ and $d\equiv\boldsymbol{1}\; \cdot \; \mathbf{C}^{-1}_{SN} \; \cdot \;\boldsymbol{1}$, with $\boldsymbol{1}$ being the identity matrix.

\subsection{Quasars}
We use the quasar data compiled in \cite{Risaliti:2015zla} consisting on 808 objects covering the redshift range $0.06<z<6.28$. In this work the UV and X-ray fluxes are given, as well as their respective erros. In order to test our model we are interested in the distance modulus $\mu$ so we compute it using the Eq.~(5) given in the above cited work, i.e.
\be
\mu=\frac{5}{2(\gamma-1)}[\log(F_X)-\gamma\log(F_{UV})-\beta']
\ee
where $\gamma=0.6$ is the average value of the free parameter which relates both fluxes and $\beta'$ is an arbitrary scaling factor. As before we compute the theoretical distance modulus using Eq.~(\ref{eq:m_jla}) and marginalize over the additive constant terms so that the the final $\chi^2$ is given by Eq.~(\ref{eq:chis})
%\be
%\chi^2_{Q}=a+\log \frac{d}{2\pi}-\frac{b^2}{d},
%\ee
where now we have: $a\equiv \left(\Delta\boldsymbol{\mathcal{F}}^{Q}\right)^T \; \cdot \; \mathbf{C}^{-1}_{Q} \; \cdot \; \Delta  \boldsymbol{\mathcal{F}}^{Q}$, $b\equiv\left(\Delta \boldsymbol{\mathcal{F}}^{Q}\right)^T \; \cdot \; \mathbf{C}^{-1}_{Q} \; \cdot \;\boldsymbol{1}$ and $d\equiv\boldsymbol{1}\; \cdot \; \mathbf{C}^{-1}_{Q} \; \cdot \;\boldsymbol{1}$.

\subsection{Gamma Ray Bursts}
We consider the so-called Mayflower sample, consisting on 79 GRBs covering the redshift range $1.44<z<8.1$ \cite{Liu:2014vda}. As before we compute the theoretical distance modulus using Eq.~(\ref{eq:m_jla}) and marginalize over the constant additive term so that the the final $\chi^2$ is given by Eq.~(\ref{eq:chis})
%\be
%\chi^2_{G}=a+\log \frac{d}{2\pi}-\frac{b^2}{d},
%\ee
with $a\equiv \left(\Delta\boldsymbol{\mathcal{F}}^{G}\right)^T \; \cdot \; \mathbf{C}^{-1}_{G} \; \cdot \; \Delta  \boldsymbol{\mathcal{F}}^{G}$, $b\equiv\left(\Delta \boldsymbol{\mathcal{F}}^{G}\right)^T \; \cdot \; \mathbf{C}^{-1}_{G} \; \cdot \;\boldsymbol{1}$ and $d\equiv\boldsymbol{1}\; \cdot \; \mathbf{C}^{-1}_{G} \; \cdot \;\boldsymbol{1}$.

\subsection{Baryon Acoustic Oscillations}

The $\chi^2_{BAO}$ for Baryon Acoustic Oscillations (BAO) is defined as
\begin{equation}
\chi^2_{BAO} = \Delta \boldsymbol{\mathcal{F}}^{BAO} \; \cdot \; \mathbf{C}^{-1}_{BAO} \; \cdot \; \Delta  \boldsymbol{\mathcal{F}}^{BAO} \; ,
\end{equation}
where the quantity $\mathcal{F}^{BAO}$ can be different depending on the considered survey. We used data from the WiggleZ Dark Energy Survey, evaluated at redshifts $0.44$, $0.6$ and $0.73$, and given in Table~1 of \cite{Blake:2012pj}; in this case the quantities to be considered are the acoustic parameter
\begin{equation}\label{eq:AWiggle}
A(z) = 100  \sqrt{\Omega_{m} \, h^2} \frac{D_{V}(z)}{c \, z} \, ,
\end{equation}
and the Alcock-Paczynski distortion parameter
\begin{equation}\label{eq:FWiggle}
F(z) = (1+z)  \frac{D_{A}(z)\, H(z)}{c} \, ,
\end{equation}
where $D_{A}$ is the angular diameter distance
\begin{equation}\label{eq:dA}
D_{A}(z)  = \frac{c}{H_{0}} \frac{1}{1+z} \ \int_{0}^{z} \frac{\mathrm{d}z'}{E(z')} \; ,
\end{equation}
and $D_{V}$ is the geometric mean of the physical angular diameter distance $D_A$ and of the Hubble function $H(z)$, and defined as
\begin{equation}\label{eq:dV}
D_{V}(z)  = \left[ (1+z)^2 D^{2}_{A}(z) \frac{c \, z}{H(z,\boldsymbol{\theta})}\right]^{1/3}.
\end{equation}
We have also considered the data from the SDSS-III Baryon Oscillation Spectroscopic Survey (BOSS) DR$12$, described in \cite{bao2} and expressed as
\begin{equation}
D_{M}(z) \frac{r^{fid}_{s}(z_{d})}{r_{s}(z_{d})} \qquad \mathrm{and} \qquad H(z) \frac{r_{s}(z_{d})}{r^{fid}_{s}(z_{d})} \, ,
\end{equation}
where $r_{s}(z_{d})$ is the sound horizon evaluated at the dragging redshift $z_{d}$; and $r^{fid}_{s}(z_{d})$ is the same sound horizon but calculated for a given fiducial cosmological model used, being equal to $147.78$ Mpc \cite{bao2}. The redshift of the drag epoch is well approximated \cite{Eisenstein} by the relation
\begin{equation}\label{eq:zdrag}
z_{d} = \frac{1291 (\Omega_{m} \, h^2)^{0.251}}{1+0.659(\Omega_{m} \, h^2)^{0.828}} \left[ 1+ b_{1} (\Omega_{b} \, h^2)^{b2}\right]\; ,
\end{equation}
where
\begin{eqnarray}\label{eq:zdrag_b}
b_{1} &=& 0.313 (\Omega_{m} \, h^2)^{-0.419} \left[ 1+0.607 (\Omega_{m} \, h^2)^{0.6748}\right], \nonumber \\
b_{2} &=& 0.238 (\Omega_{m} \, h^2)^{0.223}.
\end{eqnarray}
The sound horizon  is defined as:
\begin{equation}\label{eq:soundhor}
r_{s}(z) = \int^{\infty}_{z} \frac{c_{s}(z')}{H(z')} \mathrm{d}z'\, ,
\end{equation}
with the sound speed
\begin{equation}\label{eq:soundspeed}
c_{s}(z) = \frac{c}{\sqrt{3(1+\overline{R}_{b}\, (1+z)^{-1})}} \; ,
\end{equation}
and
\begin{equation}
\overline{R}_{b} = 31500 \Omega_{b} \, h^{2} \left( T_{CMB}/ 2.7 \right)^{-4}\; ,
\end{equation}
with $T_{CMB} = 2.726$ K.
We have also considered the point $D_V(z=1.52)=3843\pm 147\frac{r_s(zd)}{r_s^{fid}(z_d)}$ Mpc \cite{Ata:2017dya} from the  extended Baryon Oscillation Spectroscopic Survey (eBOSS).
Finally we have added data points from Quasar-Lyman $\alpha$ Forest from SDSS-III BOSS DR$11$ \cite{Font-Ribera:2013wce}:
\begin{eqnarray}
\frac{D_{A}(z=2.36)}{r_{s}(z_{d})} &=& 10.8 \pm 0.4\; , \\
\frac{c}{H(z=2.36) r_{s}(z_{d})}  &=& 9.0 \pm 0.3\; .
\end{eqnarray}
To be noted that when dealing with our scenarios $2$ and $3$ in Eqs.~(\ref{eq:scenarios}), the dimensionless matter parameter $\Omega_m$ does not appear explicitly in the cosmological background $H(z)$ while it does in Eq.~(\ref{eq:AWiggle}) and Eqs.~(\ref{eq:zdrag})~-~(\ref{eq:zdrag_b}). In such equations we will consider $\Omega_m = \Omega_b + f_{\Omega_{DM}} \cdot \overline{\Omega}_{f,0}$ for case $2$ and $\Omega_m = \left( f_{\Omega_b} + f_{\Omega_{DM}}\right) \cdot \overline{\Omega}_{f,0}$ for case $3$.

\subsection{Cosmic Microwave Background data}

The $\chi^2_{CMB}$ for Cosmic Microwave Background (CMB) is defined as
\begin{equation}
\chi^2_{CMB} = \Delta \boldsymbol{\mathcal{F}}^{CMB} \; \cdot \; \mathbf{C}^{-1}_{CMB} \; \cdot \; \Delta  \boldsymbol{\mathcal{F}}^{CMB} \; ,
\end{equation}
where $\mathcal{F}^{CMB}$ is a vector of quantities taken from \cite{cmb2}, where \textit{Planck} $2015$ data release is analyzed in order to give the so-called shift parameters defined in \cite{Wang2007}. They are related to the positions of the CMB acoustic peaks which depend on the geometry of the model considered and, as such, can be used to discriminate between dark energy models of the different nature. They are defined as:
\begin{eqnarray}
R(\boldsymbol{\theta}) &\equiv& \sqrt{\Omega_m H^2_{0}} \frac{r(z_{\ast},\boldsymbol{\theta})}{c}, \nonumber \\
l_{a}(\boldsymbol{\theta}) &\equiv& \pi \frac{r(z_{\ast},\boldsymbol{\theta})}{r_{s}(z_{\ast},\boldsymbol{\theta})}\, .
\end{eqnarray}

{\renewcommand{\tabcolsep}{1.5mm}
{\renewcommand{\arraystretch}{1.5}
\begin{table*}[ht!]
\caption{Background parameters (I)}
\begin{minipage}{0.97\textwidth}
\centering
\resizebox*{\textwidth}{!}{
\begin{tabular}{c|cccccccc|cc|}
model  &$f_{\Omega_{DM}}$&$f_{\Omega_b}$ &$\Omega_k$& $h$ & $w$ & $\tilde{A}$& $\Omega_m $& $\Omega_b$&$\mathcal{B}^{i}_{\Lambda}$ & $\ln \mathcal{B}^{i}_{\Lambda}$ \\ %&  $\mathcal{B}^{i}_{\Lambda}$ & $\ln \mathcal{B}^{i}_{\Lambda}$
\hline
$\Lambda$CDM & - &-& $-0.002_{-0.002}^{+0.002}$ & $0.669_{-0.006}^{+0.006}$ & $\mathit{-1}$  & - & $0.320_{-0.006}^{+0.007}$& $0.050_{-0.001}^{+0.001}$&$1$ & $0$ \\
1&-&-&$-0.001_{-0.002}^{+0.002}$&$0.665_{-0.007}^{+0.007}$&$-0.98_{-0.01}^{+0.02}$&$3.4_{-1.3}^{+1.4}$ &$0.323_{-0.007}^{+0.007}$&$0.050_{-0.001}^{+0.001}$& $0.40$ & $-0.91$\\
2&$0.285_{-0.007}^{+0.008}$&-&$-0.004_{-0.007}^{+0.006}$&$0.69_{-0.01}^{+0.01}$&$-0.85_{-0.06}^{+0.07}$&$0.50_{-0.07}^{+0.08}$& $0.320_{-0.007}^{+0.008}$&$0.047_{-0.002}^{+0.001}$& $1.31$ & $0.27$ \\
3&$0.271_{-0.007}^{+0.008}$&$0.046_{-0.001}^{+0.001}$&$-0.003_{+0.006}^{-0.007}$&$0.69_{-0.01}^{+0.01}$&$-0.81_{-0.07}^{+0.08}$&$0.52_{-0.08}^{+0.08}$& $0.319_{-0.007}^{+0.008}$&$0.046_{-0.002}^{+0.002}$&$1.14$ & $0.13$ \\
\end{tabular}}
\label{resultados1}
\end{minipage}
\end{table*}}}

{\renewcommand{\tabcolsep}{1.5mm}
{\renewcommand{\arraystretch}{1.5}
\begin{table*}[ht!]
\caption{Background parameters (II)}
\begin{minipage}{0.54\textwidth}
\centering
\resizebox*{\textwidth}{!}{
\begin{tabular}{c|ccc|}
 &$\Omega_b h^2$&$\Omega_{DM} h^2$ &$H_0$ \\ %&  $\mathcal{B}^{i}_{\Lambda}$ & $\ln \mathcal{B}^{i}_{\Lambda}$
\hline
Planck18 $(68\%)$& $0.02240_{-0.00015}^{+0.00015}$&$0.1196_{-0.0014}^{+0.0014}$&$67.95_{-0.64}^{+0.64}$ \\
\hline
Planck18 $(95\%)$& $0.02240_{-0.00030}^{+0.00030}$&$0.1196_{-0.0027}^{+0.0027}$ &$67.9_{-1.2}^{+1.3}$\\
\hline
model 1 & $0.02231_{-0.000157}^{+0.00016}$&$0.12042_{-0.0014}^{+0.0014}$& $66.50_{-0.67}^{+0.68}$\\
\hline
model 2 &$0.02227_{-0.000167}^{+0.00016}$&$0.1286_{-0.0049}^{+0.0046}$&$68.73_{-1.27}^{+1.13}$\\
\hline
model 3 &$0.02227_{-0.00016}^{+0.00015}$&$0.1308_{-0.0046}^{+0.0045}$&$69.27_{-1.23}^{+1.21}$\\
\end{tabular}}
\label{resultados2}
\end{minipage}
\end{table*}}}

As before, $r_{s}$ is the comoving sound horizon, evaluated at the photon-decoupling redshift $z_{\ast}$, given by the fitting formula \cite{Hu1996}:
\begin{equation}{\label{eq:zdecoupl}}
z_{\ast} = 1048 \left[ 1 + 0.00124 (\Omega_{b} h^{2})^{-0.738}\right] \left(1+g_{1} (\Omega_{m} h^{2})^{g_{2}} \right) \, ,
\end{equation}
with
\begin{eqnarray}
g_{1} &=& \frac{0.0783 (\Omega_{b} h^{2})^{-0.238}}{1+39.5(\Omega_{b} h^{2})^{-0.763}}\; , \\
g_{2} &=& \frac{0.560}{1+21.1(\Omega_{b} h^{2})^{1.81}} \, ;
\end{eqnarray}
while $r$ is the comoving distance defined as:
\begin{equation}
r(z, \boldsymbol{\theta} )  = \frac{c}{H_{0}} \int_{0}^{z} \frac{\mathrm{d}z'}{E(z',\boldsymbol{\theta})} \mathrm{d}z'\; .
\end{equation}
Again, note the presence of $\Omega_m$ and $\Omega_b$ which will be both expressed as fractions of the \textit{umami} fluid in our analysis for scenarios $2$ and $3$.

\subsection{Monte Carlo Markov Chain (MCMC)}

In order to test the predictions of our theory with the available data, we implement an MCMC code \cite{Lazkoz:2010gz,Capozziello:2011tj} to minimize the total $\chi^2$ defined as
\begin{equation}
\chi^{2}= \chi_{H}^{2} + \chi_{SN}^{2}+ \chi_{Q}^{2}+ \chi_{G}^{2} + \chi_{BAO}^{2} + \chi_{CMB}^{2} \, .
\end{equation}
The applied priors span all over the range covered by the grid we have used to calculate numerically the density of the \textit{umami} fluid. Just for numerical reasons, we will define and work with the parameter $\tilde{A}\equiv-\log_{10}\overline{A}$.

Finally, in order to set up the reliability of one model against the other, we use the Bayesian Evidence, which is generally recognized as the most reliable statistical comparison tool even if it is not completely immune to problems, like its dependence on the choice of priors \cite{Nesseris:2012cq}. We calculate it using the algorithm described in \cite{Mukherjee:2005wg}; as this algorithm is stochastic, in order to take into account possible statistical noise, we run it $\sim 100$ times obtaining a distribution of values from which we extract the best value of the evidence as the median of the distribution. The Evidence, $\mathcal{E}$, is defined as the probability of the data $D$ given the model $M$ with a set of parameters $\boldsymbol{\theta}$, $\mathcal{E}(M) = \int\ \mathrm{d}\boldsymbol{\theta}\ \mathcal{L}(D|\boldsymbol{\theta},M)\ \pi(\boldsymbol{\theta}|M)$, where $\pi(\boldsymbol{\theta}|M)$ is the prior on the set of parameters, normalized to unity, and $\mathcal{L}(D|\boldsymbol{\theta},M)$ is the likelihood function.

Once the Bayesian Evidence is calculated, one can obtain the Bayes Factor, defined as the ratio of evidences of two models, $M_{i}$ and $M_{j}$, $\mathcal{B}^{i}_{j} = \mathcal{E}_{i}/\mathcal{E}_{j}$. If $\mathcal{B}^{i}_{j} > 1$,  model $M_i$ is preferred over $M_j$, given the data. We have used the $\Lambda$CDM model, separately for both values of the pivot redshift we have defined above, as reference model $M_j$.

Even if the Bayes Factor $\mathcal{B}^{i}_{j} > 1$, one is not able yet to state how much better is model $M_i$ with respect to model $M_j$. For this, we choose the widely-used Jeffreys' Scale \cite{Jeffreys98}. In general, Jeffreys' Scale states that: if $\ln \mathcal{B}^{i}_{j} < 1$, the evidence in favor of model $M_i$ is not significant; if $1 < \ln \mathcal{B}^{i}_{j} < 2.5$, the evidence is substantial; if $2.5 < \ln \mathcal{B}^{i}_{j} < 5$, is strong; if $\ln \mathcal{B}^{i}_{j} > 5$, is decisive. Negative values of $\ln \mathcal{B}^{i}_{j}$ can be easily interpreted as evidence against model $M_i$ (or in favor of model $M_j$). In \cite{Nesseris:2012cq}, it is shown that the Jeffreys' scale is not a fully-reliable tool for model comparison, but at the same time the statistical validity of the Bayes factor as an efficient model-comparison tool is not questioned: a Bayes factor $\mathcal{B}^{i}_{j}>1$ unequivocally states that the model $i$ is more likely than model $j$. We present results in both contexts for readers' interpretation.

\section{Results}\label{sec:results}

\begin{figure*}[htbp]
\begin{center}
    \includegraphics[width=0.49\textwidth]{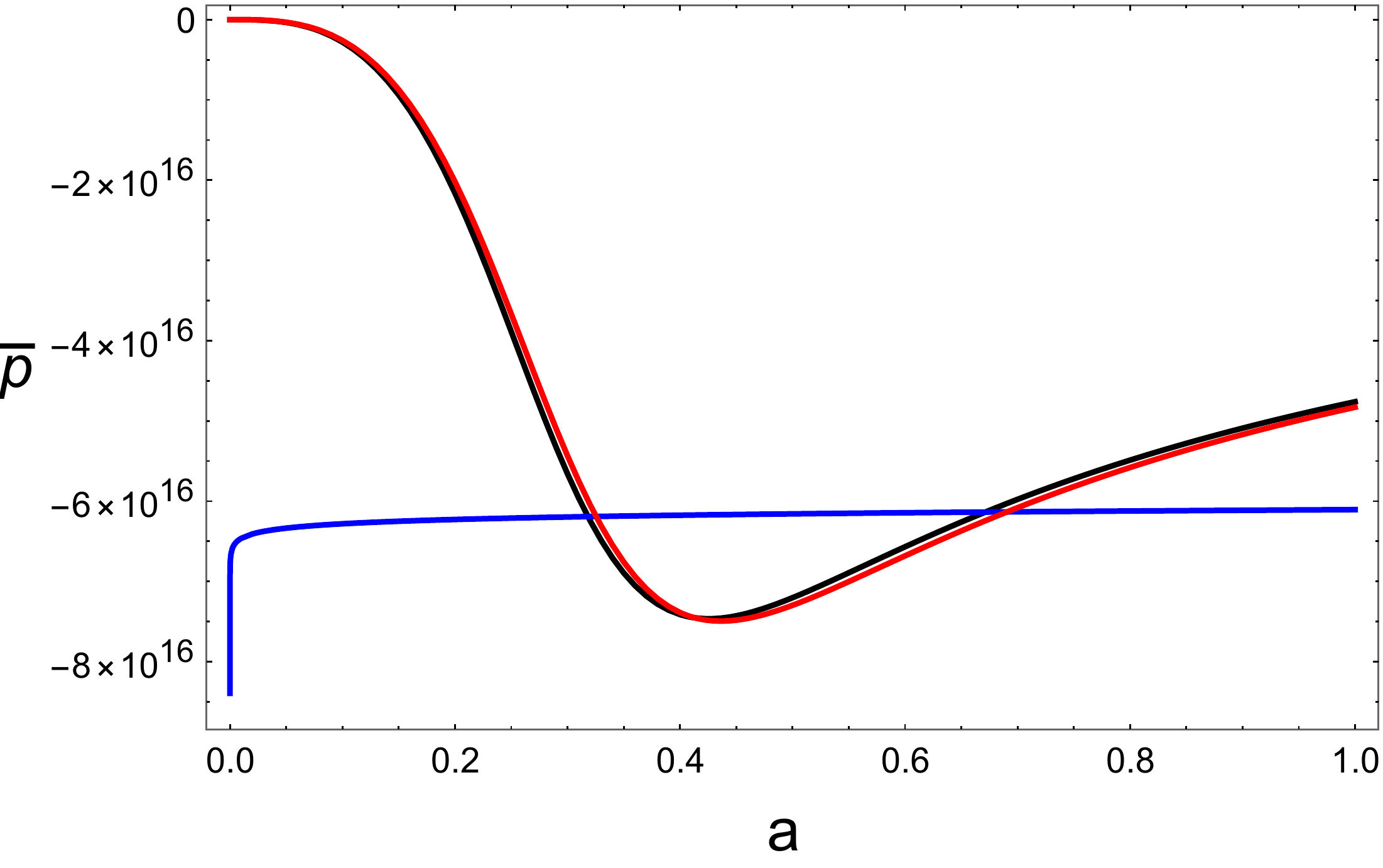}~~~
    \includegraphics[width=0.49\textwidth]{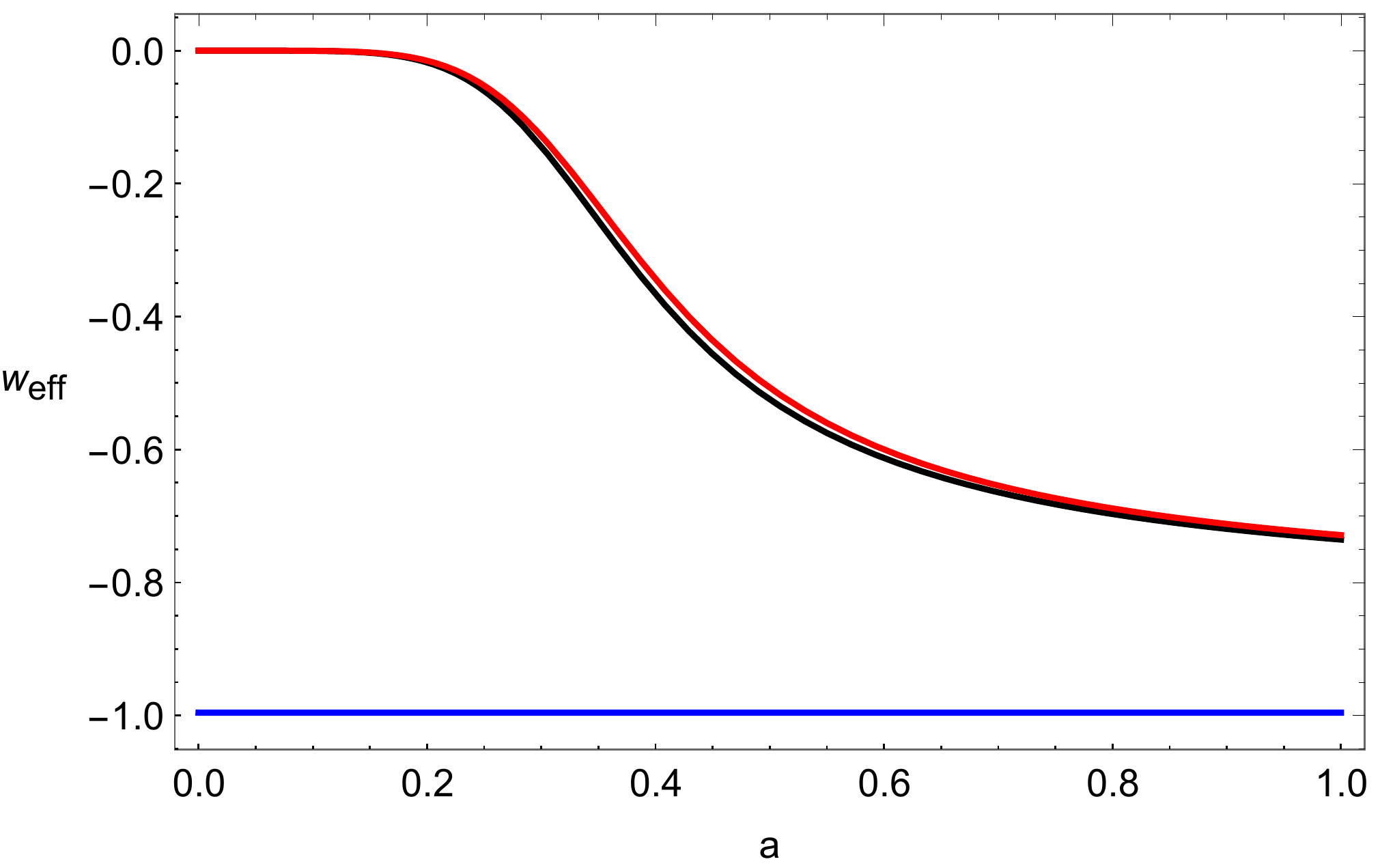}
  \caption{\textit{(Left panel.)} Evolution of the pressure with the scale factor for each of the considerate models. The blue line corresponds to model 1, where the \textit{umami} is just dark energy (DE), the black one corresponds to model 2, where the \textit{umami} is dark energy plus dark matter (DE+DM) and the red one corresponds to model 3, where the \textit{umami} is dark energy plus the total matter (DE+TM). \textit{(Right panel.)} Same as before, but for the effective equation of state parameter.}
      \label{p}
 \end{center}
\end{figure*}

In Table \ref{resultados1} we summarize the mean values of the background parameters $\{f_{\Omega_{DM}},f_{\Omega_{b}},\Omega_k,h,w,\tilde{A},\Omega_m,\Omega_b\}$ obtained for each of the scenarios considered. It can be seen that the results for all our models are in agreement with the values obtained in \textit{Planck} 2018 \cite{Aghanim:2018eyx}.

We are mostly interested in the parameters directly related with baryonic and dark matter, as they change from one model to other when considered as part of the \textit{umami} fluid. In order to make the comparison more direct, we show in Table \ref{resultados2} the value of $\Omega_b h^2$ and $\Omega_{DM}h^2$ obtained by \textit{Planck} $2018$ for the full data combination with varying $\Omega_k$\footnote{\url{https://wiki.cosmos.esa.int/planck-legacy-archive/index.php/Cosmological_Parameters}}, and by us from the MCMCs run for each one of the three scenarios we have considered. We also show the value of $H_0$, which has also a special importance due to the existing tension between the values given by local measurements \cite{Riess:2016jrr} and indirect measurements by Planck \cite{Aghanim:2016yuo}. To compute the statistics for $\Omega_{DM} h^2$ we take into account that $\Omega_{DM}=f_{\Omega_{DM}} \overline{\Omega}_{f,0}$, with $ \overline{\Omega}_{f,0}=1-\Omega_b-\Omega_r-\Omega_k$ for model 2 and $ \overline{\Omega}_{f,0}=1-\Omega_r-\Omega_k$ for model 3. Also, for the statistics of $\Omega_b h^2$, we take into account that in model three $\Omega_b=f_{\Omega_b}  \overline{\Omega}_{f,0}$.

As a general tendency we can see how the \textit{umami} fluid is generally disfavored with respect to a cosmological constant when considered as dark energy fluid only. Instead, when we consider the possibility to describe both dark matter and dark energy by one single fluid, the Bayesian Evidence becomes positive, indicating a general trend in favor of the alternative model. But it is also true that the value of the evidence is too low to set any strong preference toward it against $\Lambda$CDM. At least, it works equally good.

Another interesting point is that the value of $H_0$ increases slightly in models 2 and 3, tending \textit{to solve} the $H_0$ tension. It is also true, however, that the corresponding error also increases.
If we focus now on $\Omega_b$ and $\Omega_{m}$ we notice that they decrease in models 2 and 3. With respect to $\Omega_k$, in models 2 and 3 this parameter gets bigger but so the error does, and it is always consistent with zero.

We also want to pay attention on another interesting result of our analysis: the columns of Table \ref{resultados1} corresponding to the \textit{umami} parameters, $\tilde{A}$ and $w$. In order to analyze in detail these results and to be able to compare the models, we focus on the pressure and the effective equation of state, respectively defined as 
\begin{equation}
\overline{p}\equiv c^2 w_{eff} \overline{\Omega}_f \, , 
\end{equation}
where
\be\label{weff}
w_{eff}\equiv \frac{-1}{\displaystyle{\frac{1}{|w|}}+\frac{\overline{\Omega}_f^2}{|\overline{A}|}}.
\ee
So, as we did with the energy density, we redefine pressure normalizing over $\rho_{c,0}$ so that we have $\overline{p}\equiv\displaystyle{\frac{p}{\rho_{c,0}}}$.

Looking at the plot of $\overline{p}$ in Fig. \ref{p} one can notice how the pressure of the \textit{umami} is negative during the whole expansion history of the universe, but exhibiting different limiting behaviours at early times depending on the scenario we consider. For model 1 we see an almost constant evolution with a quite abrupt and faster decrease at early times, but still leading to a finite negative pressure. This is not really surprising, as in model $1$ the \textit{umami} fluid has been considered as a dark energy-only fluid.

For models 2 and 3, the results are very similar between them, as expected, because at the level of the background the role of the baryonic matter is somehow minor with respect to that one played by dark matter. But differently from model $1$ we now see a smooth transition from a region with negative pressure at late times, where dark energy dominates, to a region with pressure close (tending) to zero at early times, when we know that matter should be dominating the expansion of the Universe. Note that our data probe the cosmological expansions for $a>0.138$.

These different trends are related to the quantity $\tilde{A}$, which drives this transition between a matter domination epoch and a dark energy domination epoch. We notice that the obtained value is much bigger for model 1 than for models 2 and 3. To understand its role it must always be taken into account the comparison of its value with respect to $\overline{\Omega}_f$. The bump that appears in Fig. \ref{p} for model 1 occurs where the density starts to be comparable with $\tilde{A}$. For models 2 and 3 its mean value is much lower, so that its effects appear at later scale factor.

All these consideration are confirmed by the effective equation of state, $w_{eff}$. In Fig. \ref{p} we see that in the case where the \textit{umami} fluid behaves only like dark energy we do not have a clear statistical difference between the \textit{umami} and a standard $\Lambda$CDM scenario. We have some more evident changes in the cases $2$ and $3$, where the equation of state shows a clear transition on its tendency.

The leading  fact we want to remark here is that the mean values we obtain for $w_{eff}$ and $\tilde{A}$ are significantly different from what it is obtained in a model where dark energy is considered independently from the matter content, as for example $\Lambda$CDM. This means that we have a model which is drastically different from $\Lambda$CDM and that is well-fitting the data in a statistical satisfactory way.

\begin{figure*}[htbp]
\begin{center}
    \includegraphics[width=0.49\textwidth]{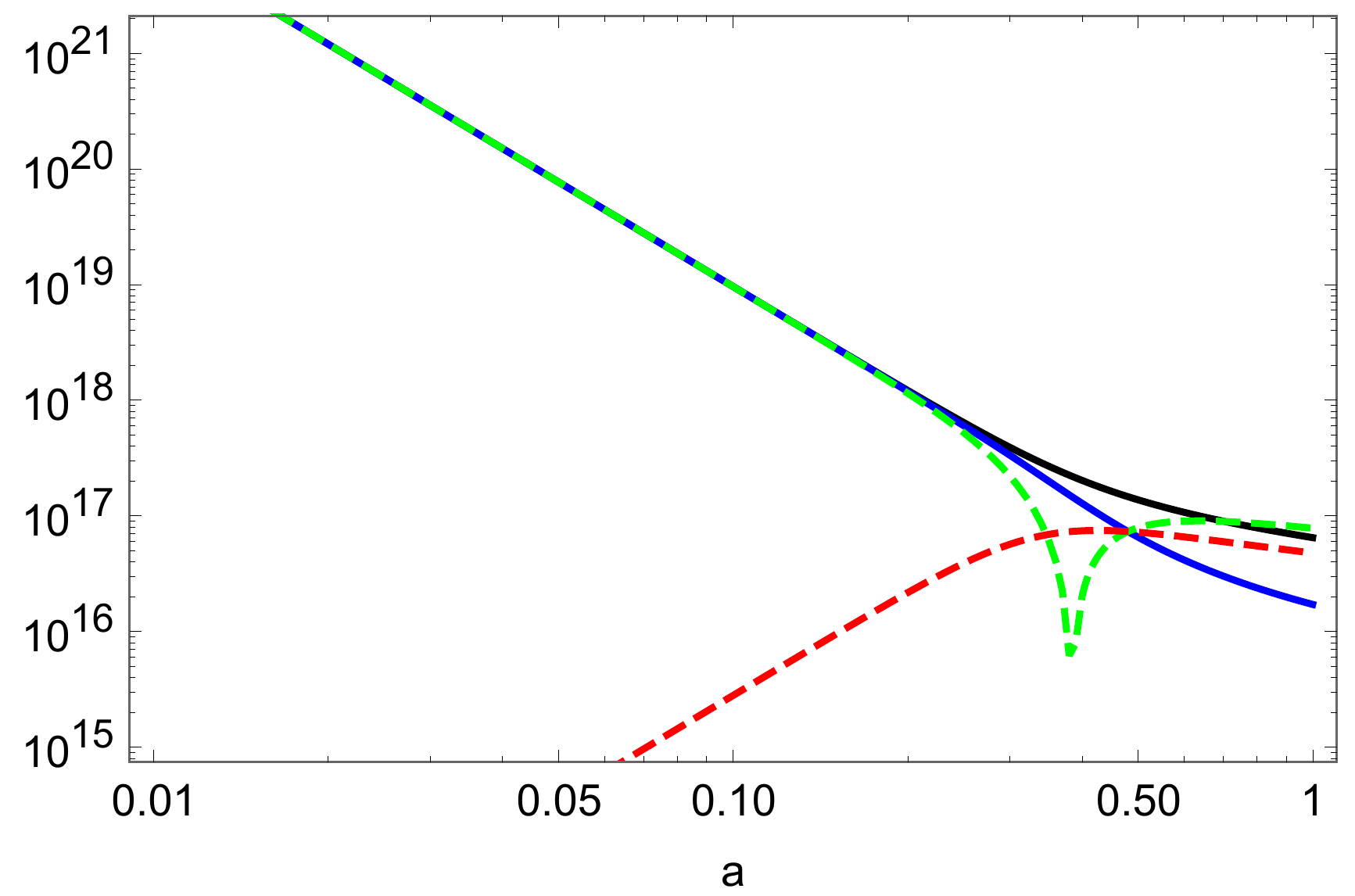}~~~
    \includegraphics[width=0.49\textwidth]{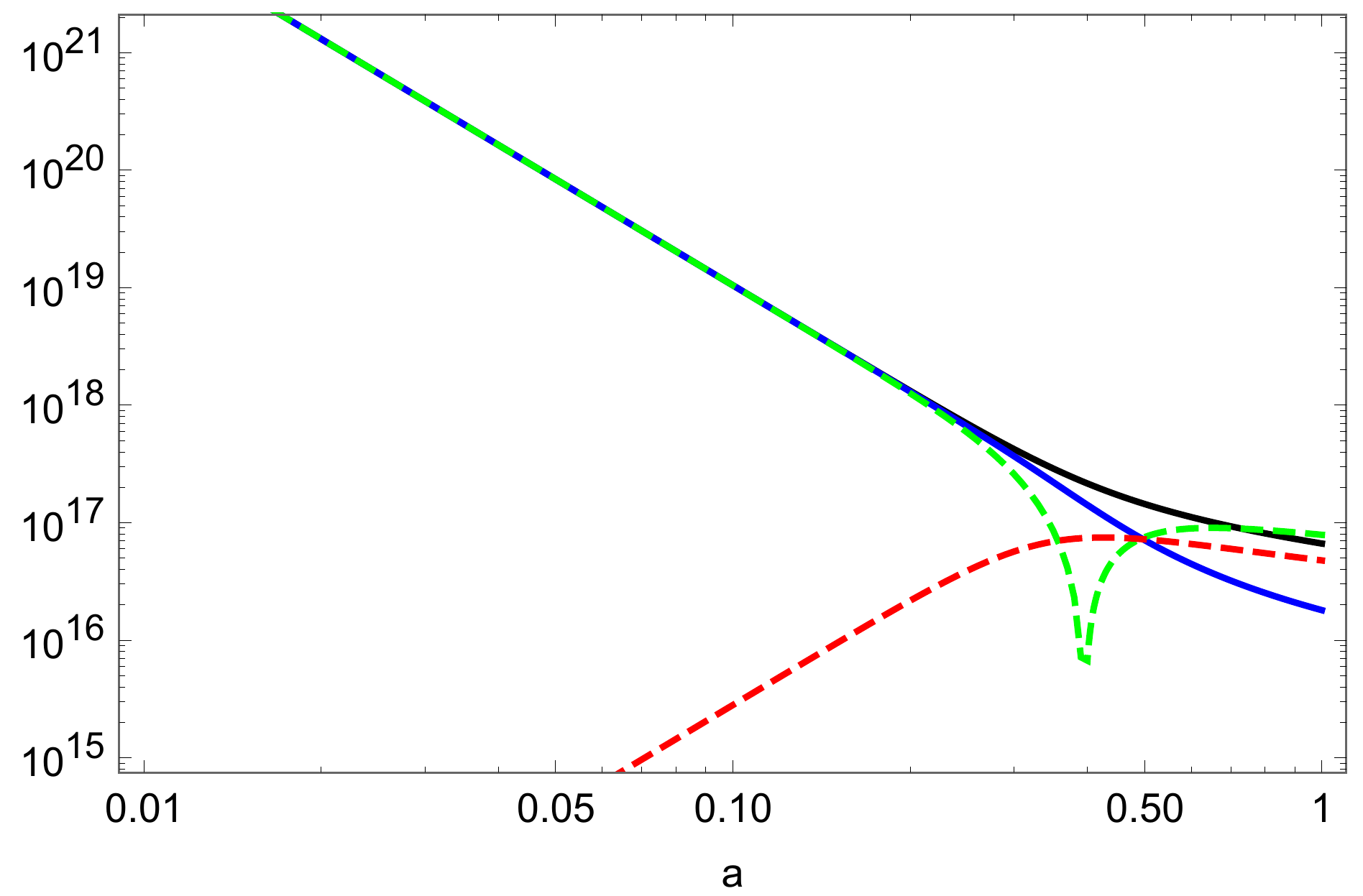}
  \caption{\textit{(Left panel.)} Energy density, pressure and energy conditions for  model 2 in absolute value. The black line corresponds to the energy density, the dashed red one corresponds to the pressure, the blue one corresponds to the WEC and the dashed green one to the SEC. \textit{(Right panel.)} Same as before, but for model 3.}
      \label{c}
 \end{center}
\end{figure*}

Finally, an interesting point to explore is to study which energy conditions are satisfied (and when) for each of the proposed \textit{umami} fluids. We have considered the energy conditions as summarized in \cite{Carroll:2003st}, and described in terms of our quantities:
\begin{itemize}
\item Null energy condition (NEC): $c^2\overline{\Omega}_f+\overline{p}\geq 0$
\item Weak energy condition (WEC): $c^2\overline{\Omega}_f+\overline{p}\geq 0$ and $\overline{\Omega}_f\geq 0$
%\item Dominant energy condition (DEC):  $c^2\overline{\Omega}_f\geq |\overline{p}|$
%\item Null Dominant energy condition (NDEC): $c^2\overline{\Omega}_f\geq |\overline{p}|$
\item Strong energy condition (SEC): $c^2\overline{\Omega}_f+\overline{p}\geq 0$ and $c^2\overline{\Omega}_f+3\overline{p}\geq 0$
\end{itemize}
We have studied these conditions for every scenario, including the cosmological constant scenario, and we have found some interesting results. First of all,  NEC and WEC conditions are perfectly satisfied by every model. As expected, $\Lambda$CDM violates the SEC condition and similarly is done by model 1, in which our \textit{umami} play the role of a fully dark energy component. On the other hand, models 2 and 3 exhibit a much more interesting evolution of the quantity corresponding to the strong energy condition. In Fig. \ref{c} we have plotted the energy density, the pressure and the WEC and SEC conditions for these two models. In both plots we see that there is cusp in the SEC condition, related to a change from negative (at large scale factor) to positive (at small scale factors) values, which means that SEC is violated when the \textit{umami} behaves as a dark energy fluid (at large scale factor), but is satisfied when its behaviours resemble that of a pressureless matter component (at low scale factor).

\section{Conclusions}\label{sec:results}

In this paper we have introduce a new phenomenological cosmological fluid with a non-linear equation of state such that it could be able to unify a dark matter cosmological behaviour with a dark energy one. We have considered three different possible contributions of such a fluid to the cosmic pie, namely: replacing only a dark energy fluid; replacing both a dark energy and a dark matter fluid; replacing both a dark energy and the total matter contribution, i.e., dark and baryonic matter. After having compared these three scenarios with the largest set of cosmological data available nowadays (as long as it is for geometrical probes; dynamical probes will be faced in a forthcoming paper), we have obtained some remarkable results we want to emphasize here. First of all we have found that our results are in agreement with the constraints given by \textit{Planck} $2018$. Secondly, for models 2 and 3, which contain a unified dark fluid and a unified baryonic-dark matter-dark energy fluid, the results obtained for the parameters $\tilde{A}$ and $w$ are completely beyond the values expected for a model with a non-unified equation of state. All this shows that this kind of unified models can properly mimic the background phenomenology even though being different from the standard cosmological $\Lambda$CDM model. Bayesian Evidence comparison shows a small but positive preference for such models. Thirdly, and not less important, although having a higher error, the mean values obtained for $H_0$ in models 2 and 3 are slightly higher than the one obtained from \textit{Planck}, and, as a consequence, closer to its local measurements. Even if this result can not be taken as conclusive, we find interesting the observed tendency to solve the $H_{0}$ tension plaguing observational cosmology today. All the observed features give us information about attractive directions to follow in the search of alternative models to $\Lambda$CDM.

\section*{Acknowledgments}
R.L. and M.O.B. were supported by the
Spanish Ministry of Economy and Competitiveness through
research projects No. FIS2014-57956-P (comprising FEDER
funds) and also by the Basque Government through research
project No. GIC17/116-IT956-16. M.O.B. acknowledges financial
support from the FPI grant BES-2015-071489.
This article is based upon work from COST Action CA15117
(CANTATA), supported by COST (European Cooperation
in Science and Technology).

\bibliography{biblio}{}

\begin{thebibliography}{62}%
\makeatletter
\providecommand \@ifxundefined [1]{%
 \@ifx{#1\undefined}
}%
\providecommand \@ifnum [1]{%
 \ifnum #1\expandafter \@firstoftwo
 \else \expandafter \@secondoftwo
 \fi
}%
\providecommand \@ifx [1]{%
 \ifx #1\expandafter \@firstoftwo
 \else \expandafter \@secondoftwo
 \fi
}%
\providecommand \natexlab [1]{#1}%
\providecommand \enquote  [1]{``#1''}%
\providecommand \bibnamefont  [1]{#1}%
\providecommand \bibfnamefont [1]{#1}%
\providecommand \citenamefont [1]{#1}%
\providecommand \href@noop [0]{\@secondoftwo}%
\providecommand \href [0]{\begingroup \@sanitize@url \@href}%
\providecommand \@href[1]{\@@startlink{#1}\@@href}%
\providecommand \@@href[1]{\endgroup#1\@@endlink}%
\providecommand \@sanitize@url [0]{\catcode `\\12\catcode `\$12\catcode
  `\&12\catcode `\#12\catcode `\^12\catcode `\_12\catcode `\%12\relax}%
\providecommand \@@startlink[1]{}%
\providecommand \@@endlink[0]{}%
\providecommand \url  [0]{\begingroup\@sanitize@url \@url }%
\providecommand \@url [1]{\endgroup\@href {#1}{\urlprefix }}%
\providecommand \urlprefix  [0]{URL }%
\providecommand \Eprint [0]{\href }%
\providecommand \doibase [0]{http://dx.doi.org/}%
\providecommand \selectlanguage [0]{\@gobble}%
\providecommand \bibinfo  [0]{\@secondoftwo}%
\providecommand \bibfield  [0]{\@secondoftwo}%
\providecommand \translation [1]{[#1]}%
\providecommand \BibitemOpen [0]{}%
\providecommand \bibitemStop [0]{}%
\providecommand \bibitemNoStop [0]{.\EOS\space}%
\providecommand \EOS [0]{\spacefactor3000\relax}%
\providecommand \BibitemShut  [1]{\csname bibitem#1\endcsname}%
\let\auto@bib@innerbib\@empty
%</preamble>
\bibitem [{\citenamefont {Riess}\ \emph {et~al.}(1998)\citenamefont {Riess}
  \emph {et~al.}}]{Riess:1998cb}%
  \BibitemOpen
  \bibfield  {author} {\bibinfo {author} {\bibfnamefont {A.~G.}\ \bibnamefont
  {Riess}} \emph {et~al.} (\bibinfo {collaboration} {Supernova Search Team}),\
  }\href@noop {} {\bibfield  {journal} {\bibinfo  {journal} {Astron. J.}\
  }\textbf {\bibinfo {volume} {116}},\ \bibinfo {pages} {1009} (\bibinfo {year}
  {1998})},\ \Eprint {http://arxiv.org/abs/astro-ph/9805201}
  {arXiv:astro-ph/9805201 [astro-ph]} \BibitemShut {NoStop}%
%%CITATION = ASTRO-PH/9805201;%%
\bibitem [{\citenamefont {Perlmutter}\ \emph {et~al.}(1999)\citenamefont
  {Perlmutter} \emph {et~al.}}]{Perlmutter:1998np}%
  \BibitemOpen
  \bibfield  {author} {\bibinfo {author} {\bibfnamefont {S.}~\bibnamefont
  {Perlmutter}} \emph {et~al.} (\bibinfo {collaboration} {Supernova Cosmology
  Project}),\ }\href@noop {} {\bibfield  {journal} {\bibinfo  {journal}
  {Astrophys. J.}\ }\textbf {\bibinfo {volume} {517}},\ \bibinfo {pages} {565}
  (\bibinfo {year} {1999})},\ \Eprint {http://arxiv.org/abs/astro-ph/9812133}
  {arXiv:astro-ph/9812133 [astro-ph]} \BibitemShut {NoStop}%
%%CITATION = ASTRO-PH/9812133;%%
\bibitem [{\citenamefont {Perlmutter}\ and\ \citenamefont
  {Schmidt}(2003)}]{Perlmutter:2003kf}%
  \BibitemOpen
  \bibfield  {author} {\bibinfo {author} {\bibfnamefont {S.}~\bibnamefont
  {Perlmutter}}\ and\ \bibinfo {author} {\bibfnamefont {B.~P.}\ \bibnamefont
  {Schmidt}},\ }\href@noop {} {\bibfield  {journal} {\bibinfo  {journal} {Lect.
  Notes Phys.}\ }\textbf {\bibinfo {volume} {598}},\ \bibinfo {pages} {195}
  (\bibinfo {year} {2003})},\ \Eprint {http://arxiv.org/abs/astro-ph/0303428}
  {arXiv:astro-ph/0303428 [astro-ph]} \BibitemShut {NoStop}%
%%CITATION = ASTRO-PH/0303428;%%
\bibitem [{\citenamefont {Aghanim}\ \emph {et~al.}(2018)\citenamefont {Aghanim}
  \emph {et~al.}}]{Aghanim:2018eyx}%
  \BibitemOpen
  \bibfield  {author} {\bibinfo {author} {\bibfnamefont {N.}~\bibnamefont
  {Aghanim}} \emph {et~al.} (\bibinfo {collaboration} {Planck}),\ }\href@noop
  {} {\  (\bibinfo {year} {2018})},\ \Eprint {http://arxiv.org/abs/1807.06209}
  {arXiv:1807.06209 [astro-ph.CO]} \BibitemShut {NoStop}%
%%CITATION = ARXIV:1807.06209;%%
\bibitem [{\citenamefont {Eisenstein}\ \emph {et~al.}(2005)\citenamefont
  {Eisenstein} \emph {et~al.}}]{sdss1}%
  \BibitemOpen
  \bibfield  {author} {\bibinfo {author} {\bibfnamefont {D.~J.}\ \bibnamefont
  {Eisenstein}} \emph {et~al.} (\bibinfo {collaboration} {SDSS}),\ }\href@noop
  {} {\bibfield  {journal} {\bibinfo  {journal} {Astrophys. J.}\ }\textbf
  {\bibinfo {volume} {633}},\ \bibinfo {pages} {560} (\bibinfo {year}
  {2005})}\BibitemShut {NoStop}%
\bibitem [{\citenamefont {Alam}\ \emph {et~al.}(2015)\citenamefont {Alam} \emph
  {et~al.}}]{sdss2}%
  \BibitemOpen
  \bibfield  {author} {\bibinfo {author} {\bibfnamefont {S.}~\bibnamefont
  {Alam}} \emph {et~al.} (\bibinfo {collaboration} {SDSS-III}),\ }\href@noop {}
  {\bibfield  {journal} {\bibinfo  {journal} {Astrophys. J. Suppl.}\ }\textbf
  {\bibinfo {volume} {219}},\ \bibinfo {pages} {12} (\bibinfo {year}
  {2015})}\BibitemShut {NoStop}%
\bibitem [{\citenamefont {Berti}\ \emph {et~al.}(2015)\citenamefont {Berti}
  \emph {et~al.}}]{Berti:2015itd}%
  \BibitemOpen
  \bibfield  {author} {\bibinfo {author} {\bibfnamefont {E.}~\bibnamefont
  {Berti}} \emph {et~al.},\ }\href@noop {} {\bibfield  {journal} {\bibinfo
  {journal} {Class. Quant. Grav.}\ }\textbf {\bibinfo {volume} {32}},\ \bibinfo
  {pages} {243001} (\bibinfo {year} {2015})}\BibitemShut {NoStop}%
\bibitem [{\citenamefont {Clifton}\ \emph {et~al.}(2012)\citenamefont
  {Clifton}, \citenamefont {Ferreira}, \citenamefont {Padilla},\ and\
  \citenamefont {Skordis}}]{Review}%
  \BibitemOpen
  \bibfield  {author} {\bibinfo {author} {\bibfnamefont {T.}~\bibnamefont
  {Clifton}}, \bibinfo {author} {\bibfnamefont {P.~G.}\ \bibnamefont
  {Ferreira}}, \bibinfo {author} {\bibfnamefont {A.}~\bibnamefont {Padilla}}, \
  and\ \bibinfo {author} {\bibfnamefont {C.}~\bibnamefont {Skordis}},\
  }\href@noop {} {\bibfield  {journal} {\bibinfo  {journal} {Phys. Rept.}\
  }\textbf {\bibinfo {volume} {513}},\ \bibinfo {pages} {1} (\bibinfo {year}
  {2012})}\BibitemShut {NoStop}%
\bibitem [{\citenamefont {Frieman}\ \emph {et~al.}(2008)\citenamefont
  {Frieman}, \citenamefont {Turner},\ and\ \citenamefont
  {Huterer}}]{Frieman:2008sn}%
  \BibitemOpen
  \bibfield  {author} {\bibinfo {author} {\bibfnamefont {J.}~\bibnamefont
  {Frieman}}, \bibinfo {author} {\bibfnamefont {M.}~\bibnamefont {Turner}}, \
  and\ \bibinfo {author} {\bibfnamefont {D.}~\bibnamefont {Huterer}},\
  }\href@noop {} {\bibfield  {journal} {\bibinfo  {journal} {Ann. Rev. Astron.
  Astrophys.}\ }\textbf {\bibinfo {volume} {46}},\ \bibinfo {pages} {385}
  (\bibinfo {year} {2008})}\BibitemShut {NoStop}%
\bibitem [{\citenamefont {Li}\ \emph {et~al.}(2011)\citenamefont {Li},
  \citenamefont {Li}, \citenamefont {Wang},\ and\ \citenamefont
  {Wang}}]{Li:2011sd}%
  \BibitemOpen
  \bibfield  {author} {\bibinfo {author} {\bibfnamefont {M.}~\bibnamefont
  {Li}}, \bibinfo {author} {\bibfnamefont {X.-D.}\ \bibnamefont {Li}}, \bibinfo
  {author} {\bibfnamefont {S.}~\bibnamefont {Wang}}, \ and\ \bibinfo {author}
  {\bibfnamefont {Y.}~\bibnamefont {Wang}},\ }\href@noop {} {\bibfield
  {journal} {\bibinfo  {journal} {Commun. Theor. Phys.}\ }\textbf {\bibinfo
  {volume} {56}},\ \bibinfo {pages} {525} (\bibinfo {year} {2011})}\BibitemShut
  {NoStop}%
\bibitem [{\citenamefont {Cooray}\ and\ \citenamefont
  {Huterer}(1999)}]{Cooray:1999da}%
  \BibitemOpen
  \bibfield  {author} {\bibinfo {author} {\bibfnamefont {A.~R.}\ \bibnamefont
  {Cooray}}\ and\ \bibinfo {author} {\bibfnamefont {D.}~\bibnamefont
  {Huterer}},\ }\href@noop {} {\bibfield  {journal} {\bibinfo  {journal}
  {Astrophys. J.}\ }\textbf {\bibinfo {volume} {513}},\ \bibinfo {pages} {L95}
  (\bibinfo {year} {1999})}\BibitemShut {NoStop}%
\bibitem [{\citenamefont {Efstathiou}(1999)}]{Efstathiou:1999tm}%
  \BibitemOpen
  \bibfield  {author} {\bibinfo {author} {\bibfnamefont {G.}~\bibnamefont
  {Efstathiou}},\ }\href@noop {} {\bibfield  {journal} {\bibinfo  {journal}
  {Mon. Not. Roy. Astron. Soc.}\ }\textbf {\bibinfo {volume} {310}},\ \bibinfo
  {pages} {842} (\bibinfo {year} {1999})}\BibitemShut {NoStop}%
\bibitem [{\citenamefont {Astier}(2001)}]{Astier:2000as}%
  \BibitemOpen
  \bibfield  {author} {\bibinfo {author} {\bibfnamefont {P.}~\bibnamefont
  {Astier}},\ }\href@noop {} {\bibfield  {journal} {\bibinfo  {journal} {Phys.
  Lett.}\ }\textbf {\bibinfo {volume} {B500}},\ \bibinfo {pages} {8} (\bibinfo
  {year} {2001})}\BibitemShut {NoStop}%
\bibitem [{\citenamefont {Goliath}\ \emph {et~al.}(2001)\citenamefont
  {Goliath}, \citenamefont {Amanullah}, \citenamefont {Astier}, \citenamefont
  {Goobar},\ and\ \citenamefont {Pain}}]{Goliath:2001af}%
  \BibitemOpen
  \bibfield  {author} {\bibinfo {author} {\bibfnamefont {M.}~\bibnamefont
  {Goliath}}, \bibinfo {author} {\bibfnamefont {R.}~\bibnamefont {Amanullah}},
  \bibinfo {author} {\bibfnamefont {P.}~\bibnamefont {Astier}}, \bibinfo
  {author} {\bibfnamefont {A.}~\bibnamefont {Goobar}}, \ and\ \bibinfo {author}
  {\bibfnamefont {R.}~\bibnamefont {Pain}},\ }\href@noop {} {\bibfield
  {journal} {\bibinfo  {journal} {Astron. Astrophys.}\ }\textbf {\bibinfo
  {volume} {380}},\ \bibinfo {pages} {6} (\bibinfo {year} {2001})}\BibitemShut
  {NoStop}%
\bibitem [{\citenamefont {Chevallier}\ and\ \citenamefont
  {Polarski}(2001)}]{Chevallier}%
  \BibitemOpen
  \bibfield  {author} {\bibinfo {author} {\bibfnamefont {M.}~\bibnamefont
  {Chevallier}}\ and\ \bibinfo {author} {\bibfnamefont {D.}~\bibnamefont
  {Polarski}},\ }\href@noop {} {\bibfield  {journal} {\bibinfo  {journal} {Int.
  J. Mod. Phys.}\ }\textbf {\bibinfo {volume} {D10}},\ \bibinfo {pages} {213}
  (\bibinfo {year} {2001})}\BibitemShut {NoStop}%
\bibitem [{\citenamefont {Linder}(2003)}]{Linder}%
  \BibitemOpen
  \bibfield  {author} {\bibinfo {author} {\bibfnamefont {E.~V.}\ \bibnamefont
  {Linder}},\ }\href@noop {} {\bibfield  {journal} {\bibinfo  {journal} {Phys.
  Rev. Lett.}\ }\textbf {\bibinfo {volume} {90}},\ \bibinfo {pages} {091301}
  (\bibinfo {year} {2003})}\BibitemShut {NoStop}%
\bibitem [{\citenamefont {Weller}\ and\ \citenamefont
  {Albrecht}(2002)}]{Weller:2001gf}%
  \BibitemOpen
  \bibfield  {author} {\bibinfo {author} {\bibfnamefont {J.}~\bibnamefont
  {Weller}}\ and\ \bibinfo {author} {\bibfnamefont {A.}~\bibnamefont
  {Albrecht}},\ }\href@noop {} {\bibfield  {journal} {\bibinfo  {journal}
  {Phys. Rev.}\ }\textbf {\bibinfo {volume} {D65}},\ \bibinfo {pages} {103512}
  (\bibinfo {year} {2002})}\BibitemShut {NoStop}%
\bibitem [{\citenamefont {Sahni}\ \emph {et~al.}(2003)\citenamefont {Sahni},
  \citenamefont {Saini}, \citenamefont {Starobinsky},\ and\ \citenamefont
  {Alam}}]{Sahni:2002fz}%
  \BibitemOpen
  \bibfield  {author} {\bibinfo {author} {\bibfnamefont {V.}~\bibnamefont
  {Sahni}}, \bibinfo {author} {\bibfnamefont {T.~D.}\ \bibnamefont {Saini}},
  \bibinfo {author} {\bibfnamefont {A.~A.}\ \bibnamefont {Starobinsky}}, \ and\
  \bibinfo {author} {\bibfnamefont {U.}~\bibnamefont {Alam}},\ }\href@noop {}
  {\bibfield  {journal} {\bibinfo  {journal} {JETP Lett.}\ }\textbf {\bibinfo
  {volume} {77}},\ \bibinfo {pages} {201} (\bibinfo {year} {2003})}\BibitemShut
  {NoStop}%
\bibitem [{\citenamefont {Padmanabhan}\ and\ \citenamefont
  {Choudhury}(2003)}]{Padmanabhan:2002vv}%
  \BibitemOpen
  \bibfield  {author} {\bibinfo {author} {\bibfnamefont {T.}~\bibnamefont
  {Padmanabhan}}\ and\ \bibinfo {author} {\bibfnamefont {T.~R.}\ \bibnamefont
  {Choudhury}},\ }\href@noop {} {\bibfield  {journal} {\bibinfo  {journal}
  {Mon. Not. Roy. Astron. Soc.}\ }\textbf {\bibinfo {volume} {344}},\ \bibinfo
  {pages} {823} (\bibinfo {year} {2003})}\BibitemShut {NoStop}%
\bibitem [{\citenamefont {Wetterich}(2004)}]{Wetterich:2004pv}%
  \BibitemOpen
  \bibfield  {author} {\bibinfo {author} {\bibfnamefont {C.}~\bibnamefont
  {Wetterich}},\ }\href@noop {} {\bibfield  {journal} {\bibinfo  {journal}
  {Phys. Lett.}\ }\textbf {\bibinfo {volume} {B594}},\ \bibinfo {pages} {17}
  (\bibinfo {year} {2004})}\BibitemShut {NoStop}%
\bibitem [{\citenamefont {Jassal}\ \emph {et~al.}(2005)\citenamefont {Jassal},
  \citenamefont {Bagla},\ and\ \citenamefont {Padmanabhan}}]{Jassal:2004ej}%
  \BibitemOpen
  \bibfield  {author} {\bibinfo {author} {\bibfnamefont {H.~K.}\ \bibnamefont
  {Jassal}}, \bibinfo {author} {\bibfnamefont {J.~S.}\ \bibnamefont {Bagla}}, \
  and\ \bibinfo {author} {\bibfnamefont {T.}~\bibnamefont {Padmanabhan}},\
  }\href@noop {} {\bibfield  {journal} {\bibinfo  {journal} {Mon. Not. Roy.
  Astron. Soc.}\ }\textbf {\bibinfo {volume} {356}},\ \bibinfo {pages} {L11}
  (\bibinfo {year} {2005})}\BibitemShut {NoStop}%
\bibitem [{\citenamefont {Komatsu}\ \emph {et~al.}(2009)\citenamefont {Komatsu}
  \emph {et~al.}}]{Komatsu:2008hk}%
  \BibitemOpen
  \bibfield  {author} {\bibinfo {author} {\bibfnamefont {E.}~\bibnamefont
  {Komatsu}} \emph {et~al.} (\bibinfo {collaboration} {WMAP}),\ }\href@noop {}
  {\bibfield  {journal} {\bibinfo  {journal} {Astrophys. J. Suppl.}\ }\textbf
  {\bibinfo {volume} {180}},\ \bibinfo {pages} {330} (\bibinfo {year}
  {2009})}\BibitemShut {NoStop}%
\bibitem [{\citenamefont {Wang}(2008)}]{Wang:2008vja}%
  \BibitemOpen
  \bibfield  {author} {\bibinfo {author} {\bibfnamefont {Y.}~\bibnamefont
  {Wang}},\ }\href@noop {} {\bibfield  {journal} {\bibinfo  {journal} {Phys.
  Rev.}\ }\textbf {\bibinfo {volume} {D78}},\ \bibinfo {pages} {123532}
  (\bibinfo {year} {2008})}\BibitemShut {NoStop}%
\bibitem [{\citenamefont {Ma}\ and\ \citenamefont {Zhang}(2011)}]{Ma:2011nc}%
  \BibitemOpen
  \bibfield  {author} {\bibinfo {author} {\bibfnamefont {J.-Z.}\ \bibnamefont
  {Ma}}\ and\ \bibinfo {author} {\bibfnamefont {X.}~\bibnamefont {Zhang}},\
  }\href@noop {} {\bibfield  {journal} {\bibinfo  {journal} {Phys. Lett.}\
  }\textbf {\bibinfo {volume} {B699}},\ \bibinfo {pages} {233} (\bibinfo {year}
  {2011})}\BibitemShut {NoStop}%
\bibitem [{\citenamefont {Barboza}\ \emph {et~al.}(2012)\citenamefont
  {Barboza}, \citenamefont {Santos}, \citenamefont {Costa},\ and\ \citenamefont
  {Alcaniz}}]{Barboza:2011py}%
  \BibitemOpen
  \bibfield  {author} {\bibinfo {author} {\bibfnamefont {E.~M.}\ \bibnamefont
  {Barboza}}, \bibinfo {author} {\bibfnamefont {B.}~\bibnamefont {Santos}},
  \bibinfo {author} {\bibfnamefont {F.~E.~M.}\ \bibnamefont {Costa}}, \ and\
  \bibinfo {author} {\bibfnamefont {J.~S.}\ \bibnamefont {Alcaniz}},\
  }\href@noop {} {\bibfield  {journal} {\bibinfo  {journal} {Phys. Rev.}\
  }\textbf {\bibinfo {volume} {D85}},\ \bibinfo {pages} {107304} (\bibinfo
  {year} {2012})}\BibitemShut {NoStop}%
\bibitem [{\citenamefont {Sendra}\ and\ \citenamefont
  {Lazkoz}(2012)}]{Sendra:2011pt}%
  \BibitemOpen
  \bibfield  {author} {\bibinfo {author} {\bibfnamefont {I.}~\bibnamefont
  {Sendra}}\ and\ \bibinfo {author} {\bibfnamefont {R.}~\bibnamefont
  {Lazkoz}},\ }\href@noop {} {\bibfield  {journal} {\bibinfo  {journal} {Mon.
  Not. Roy. Astron. Soc.}\ }\textbf {\bibinfo {volume} {422}},\ \bibinfo
  {pages} {776} (\bibinfo {year} {2012})}\BibitemShut {NoStop}%
\bibitem [{\citenamefont {Carroll}\ \emph {et~al.}(1992)\citenamefont
  {Carroll}, \citenamefont {Press},\ and\ \citenamefont
  {Turner}}]{Carroll:1991mt}%
  \BibitemOpen
  \bibfield  {author} {\bibinfo {author} {\bibfnamefont {S.~M.}\ \bibnamefont
  {Carroll}}, \bibinfo {author} {\bibfnamefont {W.~H.}\ \bibnamefont {Press}},
  \ and\ \bibinfo {author} {\bibfnamefont {E.~L.}\ \bibnamefont {Turner}},\
  }\href@noop {} {\bibfield  {journal} {\bibinfo  {journal} {Ann. Rev. Astron.
  Astrophys.}\ }\textbf {\bibinfo {volume} {30}},\ \bibinfo {pages} {499}
  (\bibinfo {year} {1992})}\BibitemShut {NoStop}%
\bibitem [{\citenamefont {Sahni}\ and\ \citenamefont
  {Starobinsky}(2000)}]{Sahni:1999gb}%
  \BibitemOpen
  \bibfield  {author} {\bibinfo {author} {\bibfnamefont {V.}~\bibnamefont
  {Sahni}}\ and\ \bibinfo {author} {\bibfnamefont {A.~A.}\ \bibnamefont
  {Starobinsky}},\ }\href@noop {} {\bibfield  {journal} {\bibinfo  {journal}
  {Int. J. Mod. Phys.}\ }\textbf {\bibinfo {volume} {D9}},\ \bibinfo {pages}
  {373} (\bibinfo {year} {2000})}\BibitemShut {NoStop}%
\bibitem [{\citenamefont {Peebles}\ and\ \citenamefont
  {Ratra}(2003)}]{Peebles:2002gy}%
  \BibitemOpen
  \bibfield  {author} {\bibinfo {author} {\bibfnamefont {P.~J.~E.}\
  \bibnamefont {Peebles}}\ and\ \bibinfo {author} {\bibfnamefont
  {B.}~\bibnamefont {Ratra}},\ }\href@noop {} {\bibfield  {journal} {\bibinfo
  {journal} {Rev. Mod. Phys.}\ }\textbf {\bibinfo {volume} {75}},\ \bibinfo
  {pages} {559} (\bibinfo {year} {2003})}\BibitemShut {NoStop}%
\bibitem [{\citenamefont {Ade}\ \emph {et~al.}(2016)\citenamefont {Ade} \emph
  {et~al.}}]{Planck}%
  \BibitemOpen
  \bibfield  {author} {\bibinfo {author} {\bibfnamefont {P.~A.~R.}\
  \bibnamefont {Ade}} \emph {et~al.} (\bibinfo {collaboration} {Planck}),\
  }\href@noop {} {\bibfield  {journal} {\bibinfo  {journal} {Astron.
  Astrophys.}\ }\textbf {\bibinfo {volume} {594}},\ \bibinfo {pages} {A13}
  (\bibinfo {year} {2016})}\BibitemShut {NoStop}%
\bibitem [{\citenamefont {Aubourg}\ \emph {et~al.}(2015)\citenamefont {Aubourg}
  \emph {et~al.}}]{Aubourg:2014yra}%
  \BibitemOpen
  \bibfield  {author} {\bibinfo {author} {\bibfnamefont {E.}~\bibnamefont
  {Aubourg}} \emph {et~al.},\ }\href@noop {} {\bibfield  {journal} {\bibinfo
  {journal} {Phys. Rev.}\ }\textbf {\bibinfo {volume} {D92}},\ \bibinfo {pages}
  {123516} (\bibinfo {year} {2015})}\BibitemShut {NoStop}%
\bibitem [{\citenamefont {Bull}\ \emph {et~al.}(2016)\citenamefont {Bull} \emph
  {et~al.}}]{beyond}%
  \BibitemOpen
  \bibfield  {author} {\bibinfo {author} {\bibfnamefont {P.}~\bibnamefont
  {Bull}} \emph {et~al.},\ }\href@noop {} {\bibfield  {journal} {\bibinfo
  {journal} {Phys. Dark Univ.}\ }\textbf {\bibinfo {volume} {12}},\ \bibinfo
  {pages} {56} (\bibinfo {year} {2016})}\BibitemShut {NoStop}%
\bibitem [{\citenamefont {Bernal}\ \emph {et~al.}(2016)\citenamefont {Bernal},
  \citenamefont {Verde},\ and\ \citenamefont {Riess}}]{Bernal2016}%
  \BibitemOpen
  \bibfield  {author} {\bibinfo {author} {\bibfnamefont {J.~L.}\ \bibnamefont
  {Bernal}}, \bibinfo {author} {\bibfnamefont {L.}~\bibnamefont {Verde}}, \
  and\ \bibinfo {author} {\bibfnamefont {A.~G.}\ \bibnamefont {Riess}},\
  }\href@noop {} {\bibfield  {journal} {\bibinfo  {journal} {JCAP}\ }\textbf
  {\bibinfo {volume} {1610}},\ \bibinfo {pages} {019} (\bibinfo {year}
  {2016})}\BibitemShut {NoStop}%
\bibitem [{\citenamefont {Kamenshchik}\ \emph {et~al.}(2001)\citenamefont
  {Kamenshchik}, \citenamefont {Moschella},\ and\ \citenamefont
  {Pasquier}}]{Kamenshchik:2001cp}%
  \BibitemOpen
  \bibfield  {author} {\bibinfo {author} {\bibfnamefont {A.~{\relax Yu}.}\
  \bibnamefont {Kamenshchik}}, \bibinfo {author} {\bibfnamefont
  {U.}~\bibnamefont {Moschella}}, \ and\ \bibinfo {author} {\bibfnamefont
  {V.}~\bibnamefont {Pasquier}},\ }\href@noop {} {\bibfield  {journal}
  {\bibinfo  {journal} {Phys. Lett.}\ }\textbf {\bibinfo {volume} {B511}},\
  \bibinfo {pages} {265} (\bibinfo {year} {2001})},\ \Eprint
  {http://arxiv.org/abs/gr-qc/0103004} {arXiv:gr-qc/0103004 [gr-qc]}
  \BibitemShut {NoStop}%
%%CITATION = GR-QC/0103004;%%
\bibitem [{\citenamefont {De-Santiago}\ \emph {et~al.}(2012)\citenamefont
  {De-Santiago}, \citenamefont {Wands},\ and\ \citenamefont
  {Wang}}]{DeSantiago:2012xh}%
  \BibitemOpen
  \bibfield  {author} {\bibinfo {author} {\bibfnamefont {J.}~\bibnamefont
  {De-Santiago}}, \bibinfo {author} {\bibfnamefont {D.}~\bibnamefont {Wands}},
  \ and\ \bibinfo {author} {\bibfnamefont {Y.}~\bibnamefont {Wang}},\ }in\
  \href@noop {} {\emph {\bibinfo {booktitle} {{6th International Meeting on
  Gravitation and Cosmology Guadalajara, Jalisco, Mexico, May 21-25, 2012}}}}\
  (\bibinfo {year} {2012})\ \Eprint {http://arxiv.org/abs/1209.0563}
  {arXiv:1209.0563 [astro-ph.CO]} \BibitemShut {NoStop}%
%%CITATION = ARXIV:1209.0563;%%
\bibitem [{\citenamefont {Pourhassan}\ and\ \citenamefont
  {Kahya}(2014)}]{Pourhassan:2014ika}%
  \BibitemOpen
  \bibfield  {author} {\bibinfo {author} {\bibfnamefont {B.}~\bibnamefont
  {Pourhassan}}\ and\ \bibinfo {author} {\bibfnamefont {E.~O.}\ \bibnamefont
  {Kahya}},\ }\href@noop {} {\bibfield  {journal} {\bibinfo  {journal} {Adv.
  High Energy Phys.}\ }\textbf {\bibinfo {volume} {2014}},\ \bibinfo {pages}
  {231452} (\bibinfo {year} {2014})},\ \Eprint {http://arxiv.org/abs/1405.0667}
  {arXiv:1405.0667 [gr-qc]} \BibitemShut {NoStop}%
%%CITATION = ARXIV:1405.0667;%%
\bibitem [{\citenamefont {Berteaud}\ \emph {et~al.}(2018)\citenamefont
  {Berteaud}, \citenamefont {Pasquet}, \citenamefont {Schücker},\ and\
  \citenamefont {Tilquin}}]{Berteaud:2018ifl}%
  \BibitemOpen
  \bibfield  {author} {\bibinfo {author} {\bibfnamefont {J.}~\bibnamefont
  {Berteaud}}, \bibinfo {author} {\bibfnamefont {J.}~\bibnamefont {Pasquet}},
  \bibinfo {author} {\bibfnamefont {T.}~\bibnamefont {Schücker}}, \ and\
  \bibinfo {author} {\bibfnamefont {A.}~\bibnamefont {Tilquin}},\ }\href@noop
  {} {\  (\bibinfo {year} {2018})},\ \Eprint {http://arxiv.org/abs/1807.05068}
  {arXiv:1807.05068 [gr-qc]} \BibitemShut {NoStop}%
%%CITATION = ARXIV:1807.05068;%%
\bibitem [{\citenamefont {Ananda}\ and\ \citenamefont
  {Bruni}(2006)}]{Ananda:2005xp}%
  \BibitemOpen
  \bibfield  {author} {\bibinfo {author} {\bibfnamefont {K.~N.}\ \bibnamefont
  {Ananda}}\ and\ \bibinfo {author} {\bibfnamefont {M.}~\bibnamefont {Bruni}},\
  }\href@noop {} {\bibfield  {journal} {\bibinfo  {journal} {Phys. Rev.}\
  }\textbf {\bibinfo {volume} {D74}},\ \bibinfo {pages} {023523} (\bibinfo
  {year} {2006})},\ \Eprint {http://arxiv.org/abs/astro-ph/0512224}
  {arXiv:astro-ph/0512224 [astro-ph]} \BibitemShut {NoStop}%
%%CITATION = ASTRO-PH/0512224;%%
\bibitem [{\citenamefont {Dutta}\ \emph {et~al.}(2010)\citenamefont {Dutta},
  \citenamefont {Chakraborty},\ and\ \citenamefont {Ansari}}]{Dutta:2010yu}%
  \BibitemOpen
  \bibfield  {author} {\bibinfo {author} {\bibfnamefont {J.}~\bibnamefont
  {Dutta}}, \bibinfo {author} {\bibfnamefont {S.}~\bibnamefont {Chakraborty}},
  \ and\ \bibinfo {author} {\bibfnamefont {M.}~\bibnamefont {Ansari}},\
  }\href@noop {} {\bibfield  {journal} {\bibinfo  {journal} {Int. J. Theor.
  Phys.}\ }\textbf {\bibinfo {volume} {49}},\ \bibinfo {pages} {2680} (\bibinfo
  {year} {2010})},\ \Eprint {http://arxiv.org/abs/1006.2206} {arXiv:1006.2206
  [gr-qc]} \BibitemShut {NoStop}%
%%CITATION = ARXIV:1006.2206;%%
\bibitem [{\citenamefont {Chen}\ \emph {et~al.}(2015)\citenamefont {Chen},
  \citenamefont {Gibbons},\ and\ \citenamefont {Yang}}]{Chen:2015kza}%
  \BibitemOpen
  \bibfield  {author} {\bibinfo {author} {\bibfnamefont {S.}~\bibnamefont
  {Chen}}, \bibinfo {author} {\bibfnamefont {G.~W.}\ \bibnamefont {Gibbons}}, \
  and\ \bibinfo {author} {\bibfnamefont {Y.}~\bibnamefont {Yang}},\ }\href@noop
  {} {\bibfield  {journal} {\bibinfo  {journal} {JCAP}\ }\textbf {\bibinfo
  {volume} {1505}},\ \bibinfo {pages} {020} (\bibinfo {year} {2015})},\ \Eprint
  {http://arxiv.org/abs/1502.05042} {arXiv:1502.05042 [gr-qc]} \BibitemShut
  {NoStop}%
%%CITATION = ARXIV:1502.05042;%%
\bibitem [{\citenamefont {Dabrowski}(2014)}]{Dabrowski:2014fha}%
  \BibitemOpen
  \bibfield  {author} {\bibinfo {author} {\bibfnamefont {M.~P.}\ \bibnamefont
  {Dabrowski}},\ }in\ \href@noop {} {\emph {\bibinfo {booktitle} {Mathematical
  Structures of the Universe}}},\ \bibinfo {editor} {edited by\ \bibinfo
  {editor} {\bibfnamefont {M.}~\bibnamefont {Heller}}, \bibinfo {editor}
  {\bibfnamefont {M.}~\bibnamefont {Eckstein}}, \ and\ \bibinfo {editor}
  {\bibfnamefont {S.}~\bibnamefont {Fast}}}\ (\bibinfo {year} {2014})\ pp.\
  \bibinfo {pages} {101--118},\ \Eprint {http://arxiv.org/abs/1407.4851}
  {arXiv:1407.4851 [gr-qc]} \BibitemShut {NoStop}%
%%CITATION = ARXIV:1407.4851;%%
\bibitem [{\citenamefont {Moresco}(2015)}]{Moresco15}%
  \BibitemOpen
  \bibfield  {author} {\bibinfo {author} {\bibfnamefont {M.}~\bibnamefont
  {Moresco}},\ }\href@noop {} {\bibfield  {journal} {\bibinfo  {journal} {Mon.
  Not. Roy. Astron. Soc.}\ }\textbf {\bibinfo {volume} {450}},\ \bibinfo
  {pages} {L16} (\bibinfo {year} {2015})}\BibitemShut {NoStop}%
\bibitem [{\citenamefont {Scolnic}\ \emph {et~al.}(2018)\citenamefont {Scolnic}
  \emph {et~al.}}]{Scolnic:2017caz}%
  \BibitemOpen
  \bibfield  {author} {\bibinfo {author} {\bibfnamefont {D.~M.}\ \bibnamefont
  {Scolnic}} \emph {et~al.},\ }\href@noop {} {\bibfield  {journal} {\bibinfo
  {journal} {Astrophys. J.}\ }\textbf {\bibinfo {volume} {859}},\ \bibinfo
  {pages} {101} (\bibinfo {year} {2018})},\ \Eprint
  {http://arxiv.org/abs/1710.00845} {arXiv:1710.00845 [astro-ph.CO]}
  \BibitemShut {NoStop}%
%%CITATION = ARXIV:1710.00845;%%
\bibitem [{\citenamefont {Conley}\ \emph {et~al.}(2011)\citenamefont {Conley},
  \citenamefont {Ellis},\ and\ \citenamefont {Neill}}]{conley}%
  \BibitemOpen
  \bibfield  {author} {\bibinfo {author} {\bibfnamefont {A.}~\bibnamefont
  {Conley}}, \bibinfo {author} {\bibfnamefont {R.~S.}\ \bibnamefont {Ellis}}, \
  and\ \bibinfo {author} {\bibfnamefont {J.~D.}\ \bibnamefont {Neill}},\
  }\href@noop {} {\bibfield  {journal} {\bibinfo  {journal} {ApJS}\ }\textbf
  {\bibinfo {volume} {192 1}} (\bibinfo {year} {2011})}\BibitemShut {NoStop}%
\bibitem [{\citenamefont {Risaliti}\ and\ \citenamefont
  {Lusso}(2015)}]{Risaliti:2015zla}%
  \BibitemOpen
  \bibfield  {author} {\bibinfo {author} {\bibfnamefont {G.}~\bibnamefont
  {Risaliti}}\ and\ \bibinfo {author} {\bibfnamefont {E.}~\bibnamefont
  {Lusso}},\ }\href@noop {} {\bibfield  {journal} {\bibinfo  {journal}
  {Astrophys. J.}\ }\textbf {\bibinfo {volume} {815}},\ \bibinfo {pages} {33}
  (\bibinfo {year} {2015})},\ \Eprint {http://arxiv.org/abs/1505.07118}
  {arXiv:1505.07118 [astro-ph.CO]} \BibitemShut {NoStop}%
%%CITATION = ARXIV:1505.07118;%%
\bibitem [{\citenamefont {Liu}\ and\ \citenamefont {Wei}(2015)}]{Liu:2014vda}%
  \BibitemOpen
  \bibfield  {author} {\bibinfo {author} {\bibfnamefont {J.}~\bibnamefont
  {Liu}}\ and\ \bibinfo {author} {\bibfnamefont {H.}~\bibnamefont {Wei}},\
  }\href@noop {} {\bibfield  {journal} {\bibinfo  {journal} {Gen. Rel. Grav.}\
  }\textbf {\bibinfo {volume} {47}},\ \bibinfo {pages} {141} (\bibinfo {year}
  {2015})},\ \Eprint {http://arxiv.org/abs/1410.3960} {arXiv:1410.3960
  [astro-ph.CO]} \BibitemShut {NoStop}%
%%CITATION = ARXIV:1410.3960;%%
\bibitem [{\citenamefont {Blake}\ \emph {et~al.}(2012)\citenamefont {Blake}
  \emph {et~al.}}]{Blake:2012pj}%
  \BibitemOpen
  \bibfield  {author} {\bibinfo {author} {\bibfnamefont {C.}~\bibnamefont
  {Blake}} \emph {et~al.},\ }\href@noop {} {\bibfield  {journal} {\bibinfo
  {journal} {Mon. Not. Roy. Astron. Soc.}\ }\textbf {\bibinfo {volume} {425}},\
  \bibinfo {pages} {405} (\bibinfo {year} {2012})}\BibitemShut {NoStop}%
\bibitem [{\citenamefont {Alam}\ \emph {et~al.}(2017)\citenamefont {Alam} \emph
  {et~al.}}]{bao2}%
  \BibitemOpen
  \bibfield  {author} {\bibinfo {author} {\bibfnamefont {S.}~\bibnamefont
  {Alam}} \emph {et~al.} (\bibinfo {collaboration} {BOSS}),\ }\href@noop {}
  {\bibfield  {journal} {\bibinfo  {journal} {Mon. Not. Roy. Astron. Soc.}\
  }\textbf {\bibinfo {volume} {470}},\ \bibinfo {pages} {2617} (\bibinfo {year}
  {2017})},\ \Eprint {http://arxiv.org/abs/1607.03155} {arXiv:1607.03155
  [astro-ph.CO]} \BibitemShut {NoStop}%
%%CITATION = ARXIV:1607.03155;%%
\bibitem [{\citenamefont {Eisenstein}\ and\ \citenamefont
  {Hu}(1998)}]{Eisenstein}%
  \BibitemOpen
  \bibfield  {author} {\bibinfo {author} {\bibfnamefont {D.~J.}\ \bibnamefont
  {Eisenstein}}\ and\ \bibinfo {author} {\bibfnamefont {W.}~\bibnamefont
  {Hu}},\ }\href@noop {} {\bibfield  {journal} {\bibinfo  {journal} {Astrophys.
  J.}\ }\textbf {\bibinfo {volume} {496}},\ \bibinfo {pages} {605} (\bibinfo
  {year} {1998})}\BibitemShut {NoStop}%
\bibitem [{\citenamefont {Ata}\ \emph {et~al.}(2018)\citenamefont {Ata} \emph
  {et~al.}}]{Ata:2017dya}%
  \BibitemOpen
  \bibfield  {author} {\bibinfo {author} {\bibfnamefont {M.}~\bibnamefont
  {Ata}} \emph {et~al.},\ }\href@noop {} {\bibfield  {journal} {\bibinfo
  {journal} {Mon. Not. Roy. Astron. Soc.}\ }\textbf {\bibinfo {volume} {473}},\
  \bibinfo {pages} {4773} (\bibinfo {year} {2018})},\ \Eprint
  {http://arxiv.org/abs/1705.06373} {arXiv:1705.06373 [astro-ph.CO]}
  \BibitemShut {NoStop}%
%%CITATION = ARXIV:1705.06373;%%
\bibitem [{\citenamefont {Font-Ribera}\ \emph {et~al.}(2014)\citenamefont
  {Font-Ribera} \emph {et~al.}}]{Font-Ribera:2013wce}%
  \BibitemOpen
  \bibfield  {author} {\bibinfo {author} {\bibfnamefont {A.}~\bibnamefont
  {Font-Ribera}} \emph {et~al.} (\bibinfo {collaboration} {BOSS}),\ }\href@noop
  {} {\bibfield  {journal} {\bibinfo  {journal} {JCAP}\ }\textbf {\bibinfo
  {volume} {1405}},\ \bibinfo {pages} {027} (\bibinfo {year} {2014})},\ \Eprint
  {http://arxiv.org/abs/1311.1767} {arXiv:1311.1767 [astro-ph.CO]} \BibitemShut
  {NoStop}%
%%CITATION = ARXIV:1311.1767;%%
\bibitem [{\citenamefont {Wang}\ and\ \citenamefont {Dai}(2016)}]{cmb2}%
  \BibitemOpen
  \bibfield  {author} {\bibinfo {author} {\bibfnamefont {Y.}~\bibnamefont
  {Wang}}\ and\ \bibinfo {author} {\bibfnamefont {M.}~\bibnamefont {Dai}},\
  }\href@noop {} {\bibfield  {journal} {\bibinfo  {journal} {Phys. Rev.}\
  }\textbf {\bibinfo {volume} {D94}},\ \bibinfo {pages} {083521} (\bibinfo
  {year} {2016})}\BibitemShut {NoStop}%
\bibitem [{\citenamefont {Wang}\ and\ \citenamefont
  {Mukherjee}(2007)}]{Wang2007}%
  \BibitemOpen
  \bibfield  {author} {\bibinfo {author} {\bibfnamefont {Y.}~\bibnamefont
  {Wang}}\ and\ \bibinfo {author} {\bibfnamefont {P.}~\bibnamefont
  {Mukherjee}},\ }\href@noop {} {\bibfield  {journal} {\bibinfo  {journal}
  {Phys. Rev.}\ }\textbf {\bibinfo {volume} {D76}},\ \bibinfo {pages} {103533}
  (\bibinfo {year} {2007})}\BibitemShut {NoStop}%
\bibitem [{\citenamefont {Hu}\ and\ \citenamefont {Sugiyama}(1996)}]{Hu1996}%
  \BibitemOpen
  \bibfield  {author} {\bibinfo {author} {\bibfnamefont {W.}~\bibnamefont
  {Hu}}\ and\ \bibinfo {author} {\bibfnamefont {N.}~\bibnamefont {Sugiyama}},\
  }\href@noop {} {\bibfield  {journal} {\bibinfo  {journal} {Astrophys. J.}\
  }\textbf {\bibinfo {volume} {471}},\ \bibinfo {pages} {542} (\bibinfo {year}
  {1996})}\BibitemShut {NoStop}%
\bibitem [{\citenamefont {Lazkoz}\ \emph {et~al.}(2011)\citenamefont {Lazkoz},
  \citenamefont {Salzano},\ and\ \citenamefont {Sendra}}]{Lazkoz:2010gz}%
  \BibitemOpen
  \bibfield  {author} {\bibinfo {author} {\bibfnamefont {R.}~\bibnamefont
  {Lazkoz}}, \bibinfo {author} {\bibfnamefont {V.}~\bibnamefont {Salzano}}, \
  and\ \bibinfo {author} {\bibfnamefont {I.}~\bibnamefont {Sendra}},\
  }\href@noop {} {\bibfield  {journal} {\bibinfo  {journal} {Phys. Lett.}\
  }\textbf {\bibinfo {volume} {B694}},\ \bibinfo {pages} {198} (\bibinfo {year}
  {2011})},\ \Eprint {http://arxiv.org/abs/1003.6084} {arXiv:1003.6084
  [astro-ph.CO]} \BibitemShut {NoStop}%
%%CITATION = ARXIV:1003.6084;%%
\bibitem [{\citenamefont {Capozziello}\ \emph {et~al.}(2011)\citenamefont
  {Capozziello}, \citenamefont {Lazkoz},\ and\ \citenamefont
  {Salzano}}]{Capozziello:2011tj}%
  \BibitemOpen
  \bibfield  {author} {\bibinfo {author} {\bibfnamefont {S.}~\bibnamefont
  {Capozziello}}, \bibinfo {author} {\bibfnamefont {R.}~\bibnamefont {Lazkoz}},
  \ and\ \bibinfo {author} {\bibfnamefont {V.}~\bibnamefont {Salzano}},\
  }\href@noop {} {\bibfield  {journal} {\bibinfo  {journal} {Phys. Rev.}\
  }\textbf {\bibinfo {volume} {D84}},\ \bibinfo {pages} {124061} (\bibinfo
  {year} {2011})},\ \Eprint {http://arxiv.org/abs/1104.3096} {arXiv:1104.3096
  [astro-ph.CO]} \BibitemShut {NoStop}%
%%CITATION = ARXIV:1104.3096;%%
\bibitem [{\citenamefont {Nesseris}\ and\ \citenamefont
  {Garcia-Bellido}(2013)}]{Nesseris:2012cq}%
  \BibitemOpen
  \bibfield  {author} {\bibinfo {author} {\bibfnamefont {S.}~\bibnamefont
  {Nesseris}}\ and\ \bibinfo {author} {\bibfnamefont {J.}~\bibnamefont
  {Garcia-Bellido}},\ }\href@noop {} {\bibfield  {journal} {\bibinfo  {journal}
  {JCAP}\ }\textbf {\bibinfo {volume} {1308}},\ \bibinfo {pages} {036}
  (\bibinfo {year} {2013})}\BibitemShut {NoStop}%
\bibitem [{\citenamefont {Mukherjee}\ \emph {et~al.}(2006)\citenamefont
  {Mukherjee}, \citenamefont {Parkinson},\ and\ \citenamefont
  {Liddle}}]{Mukherjee:2005wg}%
  \BibitemOpen
  \bibfield  {author} {\bibinfo {author} {\bibfnamefont {P.}~\bibnamefont
  {Mukherjee}}, \bibinfo {author} {\bibfnamefont {D.}~\bibnamefont
  {Parkinson}}, \ and\ \bibinfo {author} {\bibfnamefont {A.~R.}\ \bibnamefont
  {Liddle}},\ }\href@noop {} {\bibfield  {journal} {\bibinfo  {journal}
  {Astrophys. J.}\ }\textbf {\bibinfo {volume} {638}},\ \bibinfo {pages} {L51}
  (\bibinfo {year} {2006})}\BibitemShut {NoStop}%
\bibitem [{\citenamefont {Jeffreys}(1998)}]{Jeffreys98}%
  \BibitemOpen
  \bibfield  {author} {\bibinfo {author} {\bibfnamefont {H.}~\bibnamefont
  {Jeffreys}},\ }\href@noop {} {\emph {\bibinfo {title} {The theory of
  probability}}},\ \bibinfo {edition} {3rd}\ ed.,\ \bibinfo {series} {Oxford
  Classic Texts in the Physical Sciences}, Vol.~\bibinfo {volume} {4}\
  (\bibinfo  {publisher} {Oxford University Press},\ \bibinfo {address} {The
  address},\ \bibinfo {year} {1998})\BibitemShut {NoStop}%
\bibitem [{\citenamefont {Riess}\ \emph {et~al.}(2016)\citenamefont {Riess}
  \emph {et~al.}}]{Riess:2016jrr}%
  \BibitemOpen
  \bibfield  {author} {\bibinfo {author} {\bibfnamefont {A.~G.}\ \bibnamefont
  {Riess}} \emph {et~al.},\ }\href@noop {} {\bibfield  {journal} {\bibinfo
  {journal} {Astrophys. J.}\ }\textbf {\bibinfo {volume} {826}},\ \bibinfo
  {pages} {56} (\bibinfo {year} {2016})},\ \Eprint
  {http://arxiv.org/abs/1604.01424} {arXiv:1604.01424 [astro-ph.CO]}
  \BibitemShut {NoStop}%
%%CITATION = ARXIV:1604.01424;%%
\bibitem [{\citenamefont {Aghanim}\ \emph {et~al.}(2016)\citenamefont {Aghanim}
  \emph {et~al.}}]{Aghanim:2016yuo}%
  \BibitemOpen
  \bibfield  {author} {\bibinfo {author} {\bibfnamefont {N.}~\bibnamefont
  {Aghanim}} \emph {et~al.} (\bibinfo {collaboration} {Planck}),\ }\href@noop
  {} {\bibfield  {journal} {\bibinfo  {journal} {Astron. Astrophys.}\ }\textbf
  {\bibinfo {volume} {596}},\ \bibinfo {pages} {A107} (\bibinfo {year}
  {2016})},\ \Eprint {http://arxiv.org/abs/1605.02985} {arXiv:1605.02985
  [astro-ph.CO]} \BibitemShut {NoStop}%
%%CITATION = ARXIV:1605.02985;%%
\bibitem [{\citenamefont {Carroll}\ \emph {et~al.}(2003)\citenamefont
  {Carroll}, \citenamefont {Hoffman},\ and\ \citenamefont
  {Trodden}}]{Carroll:2003st}%
  \BibitemOpen
  \bibfield  {author} {\bibinfo {author} {\bibfnamefont {S.~M.}\ \bibnamefont
  {Carroll}}, \bibinfo {author} {\bibfnamefont {M.}~\bibnamefont {Hoffman}}, \
  and\ \bibinfo {author} {\bibfnamefont {M.}~\bibnamefont {Trodden}},\
  }\href@noop {} {\bibfield  {journal} {\bibinfo  {journal} {Phys. Rev.}\
  }\textbf {\bibinfo {volume} {D68}},\ \bibinfo {pages} {023509} (\bibinfo
  {year} {2003})},\ \Eprint {http://arxiv.org/abs/astro-ph/0301273}
  {arXiv:astro-ph/0301273 [astro-ph]} \BibitemShut {NoStop}%
%%CITATION = ASTRO-PH/0301273;%%
\end{thebibliography}%

\end{document}